\documentclass[aps,pre,twocolumn,nofootinbib,tightenlines,superscriptaddress,nopacs,amsmath,amssymb,final,letterpaper]{revtex4-1}

\usepackage[utf8]{inputenc}
\usepackage{calc}
\usepackage{graphicx}
\usepackage{amsmath,amssymb,amsthm}
\usepackage{bbold}
\usepackage{bm}
\usepackage{color}
\usepackage[dvipsnames]{xcolor}
\usepackage{enumitem}
\usepackage{tikz}
\usepackage{float}
\usepackage[percent]{overpic}

\newcommand{\abs}[1]{\vert #1\vert}

\def\multiset#1#2{\ensuremath{\left(\kern-.3em\left(\genfrac{}{}{0pt}{}{#1}{#2}\right)\kern-.3em\right)}}

\usepackage[colorlinks=true,linkcolor=blue,urlcolor=blue,citecolor=blue,anchorcolor=blue]{hyperref}

\begin{document}
\title{Inference of dynamic hypergraph representations in temporal interaction data}

\author{Alec \surname{Kirkley}}
\email{alec.w.kirkley@gmail.com}
\affiliation{Institute of Data Science, University of Hong Kong, Hong Kong}
\affiliation{Department of Urban Planning and Design, University of Hong Kong, Hong Kong}
\affiliation{Urban Systems Institute, University of Hong Kong, Hong Kong}

\begin{abstract}
A range of systems across the social and natural sciences generate datasets consisting of interactions between two distinct categories of items at various instances in time. Online shopping, for example, generates purchasing events of the form (user, product, time of purchase), and mutualistic interactions in plant-pollinator systems generate pollination events of the form (insect, plant, time of pollination). These data sets can be meaningfully modeled as temporal hypergraph snapshots in which multiple items within one category (i.e. online shoppers) share a hyperedge if they interacted with a common item in the other category (i.e. purchased the same product) within a given time window, allowing for the application of hypergraph analysis techniques. However, it is often unclear how to choose the number and duration of these temporal snapshots, which have a strong influence on the final hypergraph representations. Here we propose a principled nonparametric solution to this problem by extracting temporal hypergraph snapshots that optimally capture structural regularities in temporal event data according to the minimum description length principle. We demonstrate our methods on real and synthetic datasets, finding that they can recover planted artificial hypergraph structure in the presence of considerable noise and reveal meaningful activity fluctuations in human mobility data.        
\end{abstract}

%%%%%%%%%%%%%%%%%%%%%%%%%%%%%%%%%%%%%%%%%%%%%
%%%%%%%%%%%%%%%%%%%%%%%%%%%%%%%%%%%%%%%%%%%%%

\maketitle

\section{Introduction}

The recent fast-paced development of hypergraph modeling tools has opened up many new avenues for understanding the higher order structure and dynamics of complex systems ~\cite{battiston2020networks,battiston2021physics}. Datasets arising in applications as diverse as crime prediction \cite{li2022spatial}, social media analytics \cite{proferes2021studying}, and epidemiology \cite{antelmi2020design} consist of temporal interaction events between distinct categories of items---for example, users and comment threads in social media data or infected persons and locations in epidemiological data. Such datasets can be represented as temporal hypergraph snapshots, allowing for the application of centrality measures, community detection methods, link prediction algorithms, and dynamical models specifically tailored for capturing the structure and dynamics of higher order interactions. These higher order interactions produce emergent behaviors not present in traditional network representations~\cite{battiston2020networks,battiston2021physics}.In the hypergraph representation, items of one category (e.g. social media users) are represented as nodes and share a hyperedge if they were each involved in an interaction event with the same item of the other category (e.g. two users who commented on the same thread) within some specified time window.   

In some applications of interest, there is a physically meaningful time window of interest for the temporal hypergraph snapshots. For example, in epidemiology one may want to set the time scale of co-location in human mobility data to be on the order of days, in order to capture possible transmission risk from infected individuals that visited a given location. In this paper we are interested in situations where the time scale of interest is not clear ahead of time, and one must infer the characteristic time windows based on structural regularities in the event data itself. Such a need arises, for example, when identifying seasonality or anomalies in application areas without clear physical time scales such as certain online shopping behaviors~\cite{kim2015product} or vulnerabilities in  cybersystems~\cite{eren2020multi}, as well as in exploratory machine learning analyses of geolocalized events in urban planning \cite{huang2017unsupervised} and ecology \cite{balmaki2022modern}.

Existing methods for constructing networks or hypergraphs from temporal data often require each temporal event to have some non-zero duration (such representations are also called ``interval graphs'') \cite{holme2012temporal,cencetti2021temporal,myers2023topological}, but event time intervals can be hard or impossible to infer from many data generating sources, including social media checkins, online purchases, and plant-pollinator interactions. Other works choose uniform, pre-defined time windows for event aggregation \cite{lee2023temporal}, but the precise window size chosen for temporal network aggregation can have a sizeable impact on a wide variety of structural and dynamical characteristics including clustering and other centrality measures \cite{taylor2019supracentrality,amburg2020clustering}, latent node geometries \cite{huang2020temporal}, consensus dynamics \cite{neuhauser2021consensus}, controllability \cite{li2017fundamental}, epidemic spreading \cite{valdano2015analytical}, and ecological processes \cite{schwarz2020temporal}. Uniform time windows may also fail to capture the ``bursty'' dynamics of temporal network interactions, in which many events occur within short time periods that are separated by long time periods of inactivity \cite{nicosia2013graph,zha2016unfolding,cencetti2021temporal}.  

A natural way to construct hypergraph snapshots from temporal event data that overcomes these problems is to identify time windows within which the events exhibit significant shared structure. Such structural regularities can be readily identified using information theory, which allows us to quantify the level of data compression we can achieve by exploiting these regularities to transmit the data to a receiver. Among all hypergraph representations of the event data, those that better encapsulate structural regularities in the data will result in better compression from an information theoretic perspective, which can be operationalized using the Minimum Description Length (MDL) principle~\cite{rissanen1978}. The MDL principle states that the best model among a set of competing models for a given dataset is the one that can describe the data using the fewest symbols by exploiting its structural regularities~\cite{grunwald2007minimum}. The MDL principle is a powerful first-principles framework for model selection which has been employed in a range of graph mining and network science applications including community detection ~\cite{Rosvall07,Peixoto14a,Kirkley22Reps}, significant subgraph identification~\cite{koutra2014vog,wegner2014subgraph,bloem2020large,young2021hypergraph,bouritsas2021partition}, and graph summarization \cite{feng2012summarization,zhou2021dpgs,koutra2015summarizing}. 

A few existing works have examined the aggregation of temporal networks with nontrivial edge structure into representative snapshots of varying duration. In the method of Masuda and Holme~\cite{masuda2019detecting}, time is discretized into small time steps and unipartite interactions that occur within each timestep are aggregated into network snapshots. Then, a distance matrix is computed among these high resolution network snapshots using any user-specified network distance measure, and the snapshots are clustered using a hierarchical clustering algorithm to give a coarse-grained representation of the data. This method is similar in spirit to that of De Domenico et al~\cite{de2015structural}, which aggregates multilayer networks (which may or may not represent temporal snapshots) using a spectral distance between network layers. Kirkley et al~\cite{kirkley2023compressing} also approach this problem, but using a nonparametric MDL approach that is motivated by the exploitation of shared edges in these layers. These methods all differ from the one proposed in this manuscript in a few crucial ways. 

First, and most importantly, the methods in \cite{masuda2019detecting,de2015structural,kirkley2023compressing} require the fundamental measured network units (i.e. the disaggregated network layers) to have nontrivial global structure---in other words, more than just a single isolated interaction event (edge) at a given instant in time---in order to compute the network distances and clustering criteria of interest. These methods therefore require an input dataset consisting of pre-aggregated individual isolated events into coarser network snapshots to detect any signal of cohesion among the data points, which is precisely the problem studied in this paper. Second, these methods do not specifically exploit node-level structure (e.g. degrees of each node set in a bipartite representation) for compression, making them unsuitable for handling hypergraph data in which edges are shared among multiple nodes. These distinctions are critical in applications where distinct interaction events are rarely repeated. For example, in recommendation systems, a user may rarely ever consume the same product twice, which results in no meaningful shared structure among data points for the methods of \cite{masuda2019detecting,de2015structural,kirkley2023compressing} to detect. In contrast, by aiming to compress interaction event data through the exploitation of repeated items in each category (e.g. users and products) individually, the method of this manuscript can find meaningful hypergraph structure in high resolution temporal streams of interaction events. 

In this paper we first derive an objective function which computes the description length of a temporal interaction event data set under a three-part encoding that exploits structural regularities and temporal localization in the events while using a temporal hypergraph representation of the data as an intermediate step. We develop an exact polynomial time dynamic programming algorithm and a fast approximate greedy algorithm that minimize this description length objective to find the MDL-optimal configuration of temporal hypergraph snapshots associated with the event dataset. Our methods are then applied in a variety of experiments involving real and synthetic datasets to demonstrate their utility and performance. We first examine the ability for these algorithms to reconstruct planted hypergraphs in synthetic data, finding that they can recover the planted structure with high accuracy even in the presence of considerable noise. Then we apply our methods to a longitudinal location-based social network (LBSN) dataset of checkins to various locations by app users, finding that we can compress this data to automatically extract meaningful regularities in these human mobility patterns.  

%%%%%%%%%%%%%%%%%%%%%%%%%%%%%%%%%%%%%%%%%%%%%
%%%%%%%%%%%%%%%%%%%%%%%%%%%%%%%%%%%%%%%%%%%%%

\section{Methods}
\label{sec:methods}

\subsection{Temporal hypergraph binning from bipartite event data}
\label{sec:formalism}

\begin{figure*}
    \centering
    \includegraphics[width=\textwidth]{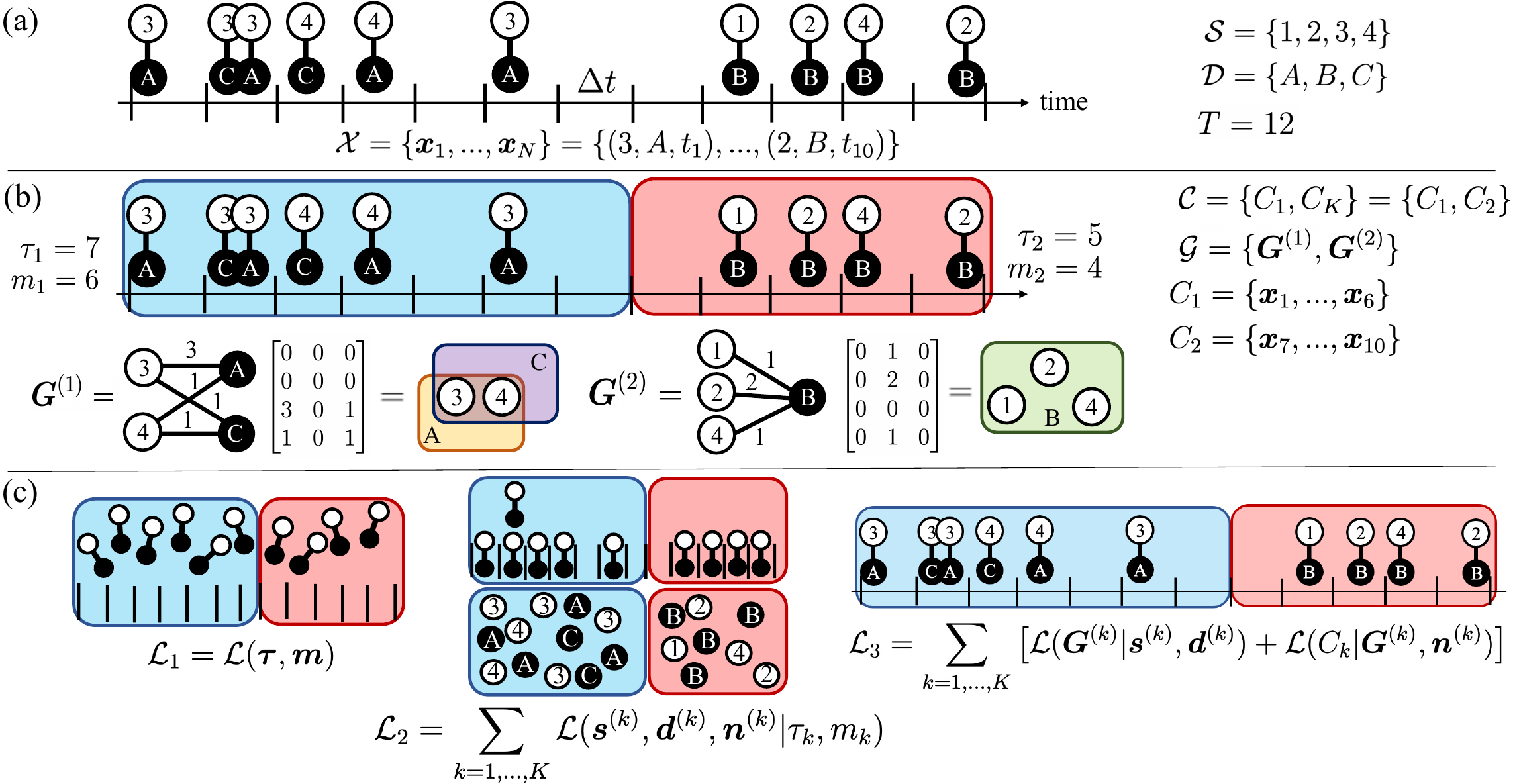}
    \caption{
    \textbf{Diagram of hypergraph binning method.} \textbf{(a)} Data set $\mathcal{X}$, consisting of $N=10$ events $(\bm{x}_i=(s_i,d_i,t_i))$ involving a ``source'' $s_i\in\mathcal{S}$ and ``destination'' $d_i\in \mathcal{D}$ interacting at time $t_i$. $\mathcal{X}$ may, for example, be used to examine co-location patterns from user-location data or co-purchasing patterns among consumers in recommendation systems analysis. Time is discretized into $T$ timesteps to allow for data compression at a desired temporal resolution $\Delta t=(t_N-t_1)/T$. \textbf{(b)} Hypergraphs $\mathcal{G}=\{\bm{G}^{(1)},\bm{G}^{(2)}\}$ extracted from partitioning the events $\mathcal{X}$ into $K=2$ clusters $\mathcal{C}=\{\{\bm{x}_1,...,\bm{x}_6\},\{\bm{x}_7,...,\bm{x}_{10}\}\}$ with localized activity patterns. The inferred weighted hypergraphs $\bm{G}^{(k)}$ are shown in both their incidence (bipartite) representation and their standard representation, with sources $s$ mapped to nodes and destinations $d$ mapped to hyperedges. \textbf{(c)} Three-stage information transmission process used to design a minimum description length objective (Eq.~\ref{eq:Ltotal}) to infer the hypergraphs $\mathcal{G}$ from event data $\mathcal{X}$. The data $\mathcal{X}$ is transmitted at increasing levels of granularity, and the optimal hypergraphs $\mathcal{G}$ (constructed using clusters $\mathcal{C}$ of events) are selected as those that minimize the description length of the transmission process.  
    }
    \label{fig:diagram}
\end{figure*}

Suppose we are given a dataset of $N$ data points (``interaction events'') $\mathcal{X}=\{\bm{x}_1,...,\bm{x}_N\}$, where each data point $\bm{x}_i=(s_i,d_i,t_i)$ consists of a source item $s_i$ of one category, a destination item $d_i$ of a different category, and a time $t_i$ when the event involving the source $s_i$ and destination $d_i$ occurred. For simplicity we can assume $\mathcal{X}$ has been ordered in time (i.e. $t_i<t_{i+1}$ for $i=1,...,N-1$), so that the entire time period of interest is $[t_1,t_N]$. We can also assume that the sources $\mathcal{S}$ and destinations $\mathcal{D}$ form disjoint sets of sizes $\abs{\mathcal{S}}=S$ and $\abs{\mathcal{D}}=D$ respectively, and that we are interested in understanding the interactions among items in only one set (e.g. $\mathcal{S}$) as mediated by the events in $\mathcal{X}$. This set comprises the nodes of the hypergraph representation we will construct. Fig.~\ref{fig:diagram}(a) shows an example of an event data set $\mathcal{X}$ consisting of $N=10$ events with sources $\mathcal{S}=\{1,2,3,4\}$, destinations $\mathcal{D}=\{A,B,C\}$, and $T=12$ time steps of size $\Delta t$ with which we discretize the event times $\{t_i\}$ (see Sec.~\ref{sec:objective} for further details).

Data in this form occurs in a wide variety of applications. Take as an example human mobility data $\mathcal{X}$, where an event $\bm{x}_i=(s_i,d_i,t_i)$ represents the presence of individual $s_i$ at location $d_i$ at time $t_i$---we will study this example in more detail in Sec.~\ref{sec:results} using location-based social network data. In this case, for applications across epidemic modelling \cite{antelmi2020design}, sociology \cite{gurukar2022leveraging}, and urban planning \cite{yu2016spatial}, one may be interested in the co-location patterns among individuals in $\mathcal{S}$. Alternatively, in recommendation systems applications, purchasing data often consists of events in which a user $s_i$ purchases a product $d_i$ at time $t_i$, and correlations among user purchasing behavior can be used for effective advertising of new products \cite{zheng2018novel}. 

A natural representation of the event data $\mathcal{X}$ in these and similar settings is as a set of hypergraph snapshots $\mathcal{G}=\{\bm{G}^{(1)},...,\bm{G}^{(K)}\}$ corresponding to consecutive non-overlapping time intervals $\{[t^{(k)}_{min},t^{(k)}_{max}]\}_{k=1}^{K}$ that partition the time interval $[t_1,t_N]$, and where node $s\in \mathcal{S}$ participates in a hyperedge labeled by $d\in \mathcal{D}$ within hypergraph $\bm{G}^{(k)}$ if and only if $s$ is involved in an event with $d$ in the time interval $[t^{(k)}_{min},t^{(k)}_{max}]$. Here we allow for a node $s$ to be repeated any number of times within a hyperedge $d$ to signal multiple events involving $\{s,d\}$ in a given time window. For maximum generality, we also allow for $\bm{G}^{(k)}$ to have self-loops (hyperedges with a single node) as well as multi-edges (distinct labelled hyperedges containing the same set of nodes). In other words, $\bm{G}^{(k)}$ is not necessarily a \emph{simple} hypergraph. The hypergraph representation $\mathcal{G}^{(k)}$ captures all the (potentially indirect) interactions among items in $\mathcal{S}$ that occur via their interactions with items in $\mathcal{D}$ during the time period $[t^{(k)}_{min},t^{(k)}_{max}]$, and can be analyzed using the wealth of newly available tools for higher order networks \cite{battiston2020networks}.

For simplicity of presentation, we will write the hypergraph snapshot $\bm{G}^{(k)}$ in its weighted bipartite (``incidence'') representation $\bm{G}^{(k)}=\big\{(s,d,G^{(k)}_{sd})\big\}_{s,d=1}^{S,D}$, where
\begin{align}\label{eq:Gdef}
G^{(k)}_{sd}=\sum_{(s_i,d_i,t_i)\in \mathcal{X}}\mathbb{1}_{t_i\in [t^{(k)}_{min},t^{(k)}_{max}]}\delta_{s_i,s}\delta_{d_i,d}    
\end{align}
is the number of events involving the node $s$ and the hyperedge $d$ within the $k$-th time window. This representation also naturally permits a symmetric treatment of the source set $\mathcal{S}$ and destination set $\mathcal{D}$, in case one is interested in the dual hypergraph representation with node set $\mathcal{D}$ and hyperedge set $\mathcal{S}$---for example, to examine similarities among products rather than users.

To construct a series of hypergraphs $\mathcal{G}=\{\bm{G}^{(k)}\}_{k=1}^{K}$ of the form in Eq.~\ref{eq:Gdef} from event data $\mathcal{X}$, one only needs to make two choices: 
\begin{enumerate}
    \item The number of temporal snapshots (``bins'') $K$.
    \item The consecutive non-overlapping time intervals $\{[t^{(k)}_{min},t^{(k)}_{max}]\}_{k=1}^{K}$. Equivalently, in discretized time (see Sec.~\ref{sec:objective}), one just needs to specify the integer-valued interval widths $\bm{\tau}=\{\tau_1,...,\tau_K\}$, where $\tau_k\Delta t=t_{max}^{(k)}-t^{(k)}_{min}$ and $\sum_{k=1}^{K}\tau_k=T$ is the total number of timesteps in our discretization.
\end{enumerate}
The integer-valued widths $\bm{\tau}$ alone fully specify the intervals $\{[t^{(k)}_{min},t^{(k)}_{max}]\}_{k=1}^{K}$ in discretized time because of the consecutive, non-overlapping nature of the intervals discussed above. For this reason, we will refer to $\bm{\tau}$ as the ``binning'' of the event data $\mathcal{X}$.

Any binning $\bm{\tau}$---a partition of (discrete) time---induces a partition of the events in $\mathcal{X}$, which we denote with $\mathcal{C}=\{C_1,...,C_K\}$. $C_k$, which we call the $k$-th ``event cluster'', is the set of data points $\bm{x}_i=(s_i,d_i,t_i)$ such that $t_i\in [t^{(k)}_{min},t^{(k)}_{max}]$. From $C_k$ we can construct the $k$-th hypergraph snapshot $\bm{G}^{(k)}$ using
\begin{align}\label{eq:GkdefCk}
G_{sd}^{(k)} = \sum_{(s_i,d_i,t_i)\in C_k}\delta_{s_i,s}\delta_{d_i,d}.  \end{align}
We denote the number of events in $C_k$ with $m_k$, and the vector of sizes for all clusters in $\mathcal{C}$ as $\bm{m}$. In this way, the integer vector $\bm{\tau}=\{\tau_1,...,\tau_K\}$ indicates the sizes of the $K$ snapshots in terms of timesteps $\Delta t$, and the integer vector $\bm{m}=\{m_1,...,m_K\}$ indicates the sizes of the snapshots in terms of the number of events $\bm{x}\in \mathcal{X}$ they contain. Note that there may be multiple binnings $\bm{\tau}$ that induce the same event partition $\mathcal{C}$, since any ``empty'' time steps $\Delta t$ (i.e., timesteps in which no events occur) at the boundary of a snapshot can be moved to an adjacent snapshot without changing the number of events occurring in each snapshot. 

In Fig.~\ref{fig:diagram}(b) we show a binning of an event dataset $\mathcal{X}$ into $K=2$ bins of widths $\bm{\tau}=\{\tau_1,\tau_2\}=\{7,5\}$, which induces an event partition $\mathcal{C}=\{\{\bm{x}_1,...,\bm{x}_6\},\{\bm{x}_7,...,\bm{x}_{10}\}\}$ and hypergraphs 
$\bm{G}^{(1)}=\{(3,A,3),(3,C,1),(4,A,1),(4,C,1)\}$, $\bm{G}^{(2)}=\{(1,B,1),(2,B,2),(4,B,1)\}$. We show each hypergraph in both its bipartite incidence representation (along with its incidence matrix), as well as in its representation with nodes in $\mathcal{S}=\{1,2,3,4\}$ and hyperedges in $\mathcal{D}=\{A,B,C\}$. 

\subsection{Minimum description length binning objective}
\label{sec:objective}

The method we present in this paper provides a principled nonparametric solution to identify hypergraph snapshots $\mathcal{G}$ of any event dataset $\mathcal{X}$ using the minimum description length (MDL) principle from information theory, which states that the best model among a set of candidate models is the one that provides the best compression (shortest description) of a dataset \cite{rissanen1978,grunwald2007minimum}. We do this by constructing a three-part encoding that allows us to gradually transmit the data $\mathcal{X}$ at increasing levels of granularity, with the hypergraphs $\mathcal{G}$ transmitted as an intermediate step in the process. The less information this transmission process requires, the more the hypergraph binning process has compressed the data $\mathcal{X}$ by capturing its statistical regularities, and the better the representation $\mathcal{G}$. The hypergraphs $\mathcal{G}$ that result in the most efficient lossless transmission of the data set $\mathcal{X}$ to a receiver (i.e., the lowest description length) give an MDL-optimal temporal hypergraph representation of $\mathcal{X}$. 

In order to construct a lossless MDL objective, we need to discretize the relevant time interval $[t_1,t_N]$ into small, uniform time steps of size $\Delta t=(t_N-t_1)/T$, where $T$ is the number of time steps. The parameter $T$ is technically a free parameter of the method to be chosen by the user, but we show empirically in Sec.~\ref{sec:results} that it has little to no impact on inference results. For this reason we consider the proposed method to be nonparametric since it has no parameters that require tuning by the user other than $T$, which can be set arbitrarily based on computational limitations (we discuss the time complexity of our methods in Sec.~\ref{sec:optimization}). Given the discretization of time into intervals of width $\Delta t$, we preprocess the data $\mathcal{X}$ by rounding each $t_i$ to the value of the closest time step, which will incur an error of at most $\Delta t/2$ for each $t_i$ and can potentially permit multiple events to occur simultaneously in the same time step. By discretizing time, we can then proceed with developing a lossless transmission scheme that results in perfect reconstruction of the discretized data $\mathcal{X}$, and whose information content is computed using discrete combinatorial structures. However, due to the rounding, we are in effect performing \emph{lossy} compression with maximum distortion $\Delta t/2$ in the time values reconstructed by a receiver. 

With this discretization in place, we can construct our MDL objective for communicating $\mathcal{X}$ using the hypergraphs $\mathcal{G}$ as an intermediate step. The fundamental mechanism behind our encoding is that we can obtain compression of event data $\mathcal{X}$ using hypergraphs $\mathcal{G}$ that are localized in time as well as with respect to sources $s$ and destinations $d$. This is made possible by our encoding exploiting the redundancies in the events $\bm{x}_i$ that take place within the event clusters $\mathcal{C}$ corresponding to these hypergraphs. This localization is also consistent with previous findings that bipartite graphs and hypergraphs display heavy-tailed (hyper-)degrees \cite{burgos2008two,lee2021hyperedges,zhang2010hypergraph}, as well as ``burstiness'' in time \cite{nicosia2013graph,zha2016unfolding,cencetti2021temporal}. 

Suppose we want to transmit the (temporally discretized) dataset $\mathcal{X}$ to a receiver. We will assume that the number of events $N$, the number of discrete timesteps $T$, the number of sources (nodes) $S$, and the number of destinations (hyperedges) $D$ are known by the receiver. These are all integer constants and are of comparatively negligible information cost to transmit, so we can safely ignore them in our formulation. Suppose now that we do not use any intermediate steps in our transmission process that exploit event redundancies, and instead choose to communicate the data $\mathcal{X}$ directly to the receiver as a set of completely independent events. The receiver knows there $N$ events $\bm{x}_i=(s_i,d_i,t_i)$, each with $S$ possible sources, $D$ possible destinations, and $T$ possible timesteps. Therefore, there are
$(SDT)^{N}$ possible configurations of the data $\mathcal{X}$, and so to specify to the receiver in binary which particular configuration corresponds to our dataset, we need to send a message of length up to approximately
\begin{align}\label{eq:L0naive}
\mathcal{L}_0=\log ((SDT)^{N}) = N\log (SDT)   
\end{align}
bits, where we've used the notation $\log\equiv \log_2$ for brevity. The quantity $\mathcal{L}_0$ is referred to as the \emph{description length} of the dataset
$\mathcal{X}$ under the na\"ive one-level encoding scheme we've devised, which only uses the global information $\{N,T,S,D\}$ to constrain the space of possible datasets $\mathcal{X}$. 

A smarter way to transmit the data $\mathcal{X}$ that exploits the redundancies we seek in our hypergraph representation $\mathcal{G}$ is to transmit $\mathcal{X}$ to the receiver in three stages by passing binary messages that communicate information at increasing levels of granularity and successively constrain the space of possible datasets $\mathcal{X}$ until there is only one remaining possibility. Each step in our transmission scheme requires an information content (e.g. description length) given by the logarithm of the number of possible message configurations, as in Eq.~\ref{eq:L0naive}. We will assume that the number of data bins (i.e. hypergraphs, event clusters) $K$ is known by the receiver, and can ignore its information along with the other constants above. Crucially, although $K$ is assumed known by the receiver, it remains a free variable in our description length optimization process, as we will see in Sec.~\ref{sec:optimization}. Our transmission process proceeds as follows:
\begin{enumerate}[leftmargin=1em]

    \item \textbf{Transmit aggregate cluster-level information (event cluster sizes $\bm{m}$ and bin widths $\bm{\tau}$):}
    \begin{itemize}
        \item To transmit $\bm{\tau}=\{\tau_k\}_{k=1}^{K}$, we must specify $K$ positive integers that sum to $T$ in a particular order---also known as a ``$K$-composition'' of $T$. There are ${T-1 \choose K-1}$ ways to do this, which can be shown using the classic ``stars and bars'' method from combinatorics \cite{feller1950}. Therefore, specifying the particular $K$-composition corresponding to $\bm{\tau}=\{\tau_k\}_{k=1}^{K}$ requires $\log {T-1 \choose K-1}$ bits of information.
        \item Similarly, $\bm{m}=\{m_k\}_{k=1}^{K}$ requires $\log {N-1 \choose K-1}$ bits of information to specify, as it consists of $K$ positive integers that sum to $N$.
    \end{itemize}
    The total information content of this first stage is therefore given by the sum of these two contributions:
    \begin{align}
    \mathcal{L}_1 = \mathcal{L}(\bm{\tau},\bm{m})  = \log {T-1 \choose K-1}+\log {N-1 \choose K-1}. 
    \end{align}

    \item \textbf{Transmit detailed cluster-level information (counts of sources, destinations, and timestamps for each event cluster $C_k$):}
    \begin{itemize}
    
        \item The number of instances (bipartite degree) of each source in event cluster $C_k$ is stored in the vector $\bm{s}^{(k)}=\{s^{(k)}_r\}_{r=1}^{S}$, with $s^{(k)}_r$ the number of occurrences of source $r$ in event cluster $C_k$. There are $\multiset{S}{m_k}$ possible ways to assign each of the $S$ sources a non-negative integer degree value such that the sum of the degrees is $m_k$, where $\multiset{y}{x}={x+y-1\choose y-1}$ is the multiset coefficient counting the number of ways to assign $x$ objects to $y$ distinct bins while allowing bins to be empty. (This result can also be found using the stars and bars argument.) Therefore, transmitting the particular counts $\bm{s}^{(k)}=\{s^{(k)}_r\}_{r=1}^{S}$ requires $\log \multiset{S}{m_k}$ bits of information. 
        
        \item The number of instances (bipartite degree) of each destination in event cluster $C_k$ is stored in the vector $\bm{d}^{(k)}=\{d^{(k)}_r\}_{r=1}^{D}$, with $d^{(k)}_r$ the number of occurrences of destination $r$ in event cluster $C_k$. Transmitting these counts requires $\log \multiset{D}{m_k}$ bits of information.

        \item The number of events $\bm{x}_i$ in event cluster $C_k$ that occur at each  discrete time step within the temporal boundaries of the cluster is stored in the vector $\bm{n}^{(k)}=\{n^{(k)}_t\}_{t=1}^{\tau_k}$. Here, $n^{(k)}_t$ the number of events within event cluster $C_k$ that fall into the $t$-th time step within the cluster's boundary (there are $\tau_k$ time steps to choose from). Transmitting these counts requires $\log \multiset{\tau_k}{m_k}$ bits of information.

    \end{itemize}
    The total information content of this second stage is therefore given by the sum of these three contributions for each cluster $C_k$:
    \begin{align}
    \mathcal{L}_2 &= \sum_{k=1}^{K}\mathcal{L}(\bm{s}^{(k)},\bm{d}^{(k)},\bm{n}^{(k)}\vert \tau_k,m_k)  \\&=\sum_{k=1}^{K}\left[\log \multiset{S}{m_k}+\log\multiset{D}{m_k}+\log\multiset{\tau_k}{m_k} \right]. 
    \end{align}

    \item \textbf{Transmit the events $\bm{x}_i=(s_i,d_i,t_i)$ within each cluster $C_k$, which fully specifies $\mathcal{X}$}:\\[0.5em]
    We have the following three constraints based on previously transmitted information
    \begin{align}
     \sum_{r=1}^{S}s^{(k)}_r &= m_k,\\
     \sum_{r=1}^{D}d^{(k)}_r &= m_k,\\
     \sum_{t=1}^{\tau_k}n^{(k)}_t &= m_k.
    \end{align}
    Therefore, the number of non-negative integer-valued 3D tensors with margins defined by $\bm{s}^{(k)},\bm{d}^{(k)},\bm{n}^{(k)}$ is the number of possibilities for $\mathcal{X}$, and the logarithm of this quantity is the information content of this last step. However, this quantity itself is difficult to compute, so we can break up this last step into two stages, one of which involves the hypergraphs we are looking for:
    \begin{itemize}
    
        \item Transmit the hypergraph $\mathcal{G}^{(k)}$ given the bipartite degree constraints $\bm{s}^{(k)},\bm{d}^{(k)}$. This requires $\log\Omega(\bm{s}^{(k)},\bm{d}^{(k)})$ bits of information, where $\Omega(\bm{s}^{(k)},\bm{d}^{(k)})$ is the number of non-negative integer-valued matrices with margins $\bm{s}^{(k)},\bm{d}^{(k)}$. $\log \Omega(\bm{s}^{(k)},\bm{d}^{(k)})$ is in general difficult to compute exactly, but can be approximated in order $\text{O}(S+D)$ time using the effective columns approximation of~\cite{jerdee2022improved}. Here we use $\text{O}(\cdot)$ to indicate ``Big O'' notation.

        \item Transmit the final data points $\bm{x}_i=(s_i,d_i,t_i)$ in cluster $C_k$ given the hypergraph $\mathcal{G}^{(k)}$ and the time step counts $\bm{n}^{(k)}$. Transmitting these requires $\log\Omega(\mathcal{G}^{(k)},\bm{n}^{(k)})$ bits of information, where $\Omega(\mathcal{G}^{(k)},\bm{n}^{(k)})$ is the number of non-negative integer-valued matrices with margins $\bm{n}^{(k)}$ and $\{G^{(k)}_{sd}\}_{s,d=1}^{S,D}$. Using the approximation in~\cite{jerdee2022improved}, $\log \Omega(\mathcal{G}^{(k)},\bm{n}^{(k)})$ can be estimated in order $\text{O}(SD+\tau_k)$ time.
        
    \end{itemize}
    The total information content of this third stage is therefore given by the sum of these two contributions for each cluster $C_k$:
    \begin{align}
    \mathcal{L}_3 &= \sum_{k=1}^{K}\big[\mathcal{L}(\bm{G}^{(k)}\vert \bm{s}^{(k)},\bm{d}^{(k)})+\mathcal{L}(C_k\vert \bm{G}^{(k)},\bm{n}^{(k)}) \big ]\\  
    &=\sum_{k=1}^{K}\big[\log \Omega(\bm{s}^{(k)},\bm{d}^{(k)})+\log\Omega(\mathcal{G}^{(k)},\bm{n}^{(k)})\big].   
    \end{align}

\end{enumerate}

Summing the description length of each stage, we have a total description length of
\begin{align}\label{eq:Ltotal}
\mathcal{L}_{total}(\mathcal{X},\bm{\tau}) = \mathcal{L}_1 + \mathcal{L}_2 + \mathcal{L}_3, 
\end{align}
where we've explicitly noted the functional dependence of $\mathcal{L}$ on the binning $\bm{\tau}$, since the description length of the data $\mathcal{X}$ under our transmission scheme---including the hypergraphs $\mathcal{G}$---can be computed when $\bm{\tau}$ is known. Fig.~\ref{fig:diagram}(c) shows a schematic of the three stages described above, for the event dataset in Fig.~\ref{fig:diagram}(a). In the next section we describe how to minimize the description length of Eq.~\ref{eq:Ltotal} over all binnings $\bm{\tau}$ to find the MDL set of hypergraphs $\mathcal{G}$ that best summarize the temporal event data $\mathcal{X}$.

We note that the transmission scheme described above and its resulting description length can also be motivated from a (microcanonical) Bayesian generative model for temporal event data, with hierarchical uniform priors corresponding to each transmission step and where description lengths are negative log-probabilities. In other words, each step in the transmission corresponds to a uniform prior over valid configurations given the constraints previously transmitted. For example, the very first transmission step corresponds to drawing $K$ bin widths $\{\tau_1,...,\tau_K\}$ uniformly at random given the constraint that they must sum to $T$. And the final transmission step corresponds to drawing the event data from a uniform distribution over all valid event configurations given the weighted hypergraph representation and temporal clustering patterns. This model---possessing conceptual similarities to a configuration model for bipartite graph data where networks are generated uniformly at random given degree constraints on both node sets---is only one possible way to generate temporal hypergraphs, but it is arguably the simplest method that can meaningfully capture statistical regularities in event data when applied to discrete time periods that contain one or zero events (which is the case for $T\to\infty$). Optimization over our objective corresponds to Maximum a Posteriori estimation in the corresponding Bayesian model, and we discuss how to do this in the next section.

\subsection{Optimization and model selection}
\label{sec:optimization}

The description length of Eq.~\ref{eq:Ltotal} amounts to a one-dimensional clustering objective over binnings $\bm{\tau}$. Therefore, if our objective consisted of independent terms for each cluster (akin to the objective for K-means clustering), then we could optimize it exactly using a dyamic programming approach ~\cite{jackson2005algorithm,bellman2013dynamic,patania2023rapid,kirkley2023compressing}. 

Only the first term in Eq.~\ref{eq:Ltotal} couples clusters together, but in the regime we are interested, we have $K\ll T,N$ and we can rewrite Eq.~\ref{eq:Ltotal} as (up to irrelevant constant factors)
\begin{align}\label{eq:Ldecoupled}
 \mathcal{L}_{total}(\mathcal{X},\bm{\tau}) = \sum_{k=1}^{K} \mathcal{L}^{(k)}_{cluster}, 
\end{align}
where 
\begin{align}\label{eq:Lk}
 \mathcal{L}^{(k)}_{cluster} &= \log (N-1)(T-1)\nonumber \\
 &+ \log \multiset{S}{m_k}\multiset{D}{m_k}\multiset{\tau_k}{m_k}  \\
 &+ \left[\log \Omega(\bm{s}^{(k)},\bm{d}^{(k)})+\log\Omega(\mathcal{G}^{(k)},\bm{n}^{(k)})\right]\nonumber.
\end{align}

We can now minimize our MDL objective in Eq.~\ref{eq:Ldecoupled} using a dynamic program. The key intuition behind this is that since the objective in Eq.~\ref{eq:Ldecoupled} is a sum of independent terms over clusters in one dimension, its minimum over the first $j$ time steps---i.e., the optimal binning $\bm{\tau}$ restricted to these first $j$ time steps---must consist of the optimal binning up to some time step $i\in \{1,...,j\}$ (excluding the $i$-th time step) plus a final cluster of time steps $i,...,j$. In other words, for all $j\in [1,T]$ we have 
\begin{align}
\mathcal{L}_{\text{MDL}}^{(j)} = \min_{i\in [1,j]}\left\{ \mathcal{L}_{\text{MDL}}^{(i-1)} + \mathcal{L}^{([i,j])}_{cluster} \right\}, 
\end{align}
where $\mathcal{L}_{\text{MDL}}^{(j)}$ is the minimum value of Eq.~\ref{eq:Ldecoupled} when we include only the first $j$ timestamps $\Delta t$, and $\mathcal{L}^{([i,j])}_{cluster}$ is the cluster-level description length of Eq.~\ref{eq:Lk} evaluated at the cluster containing consecutive time steps $\{i,...,j\}$. Setting $\mathcal{L}^{(0)}_{\text{MDL}}=0$ and iterating over all $j\in [1,T]$, we find the minimum of the description length in Eq.~\ref{eq:Ldecoupled}, giving us the optimal binning $\bm{\tau}$ for the data $\mathcal{X}$ according to the MDL principle.

In addition to finding the exact optimum over binnings $\bm{\tau}$, this approach has the advantage of automatically selecting the optimal number of bins $K$, since the entire unconstrained space of binnings $\bm{\tau}$ is explored by the algorithm. The objective function in Eq.~\ref{eq:Ldecoupled} will naturally penalize high values of $K$ since we will waste information to describe the clusters and increase the total description length if $K$ is too high. On the other hand, Eq.~\ref{eq:Ldecoupled} will also naturally penalize values of $K$ that are too low, since we will waste information describing the events within each cluster if they are too heterogeneous and/or spread over too large a time period. The MDL-optimal binning $\bm{\tau}$ therefore balances the information required to describe the clusters and the information required to describe the data within each cluster by selecting an appropriate number of clusters $K$ using the data itself. 

To quantify the extent to which our method has achieved compression over a na\"ive one-level encoding, we could take the ratio of the optimal description length $\mathcal{L}_{\text{MDL}}$ from Eq.~\ref{eq:Ldecoupled} with the description length of Eq.~\ref{eq:L0naive}. However, in our case it is of more interest to determine how much of a compression gain we achieve when we use an optimal configuration of multiple hypergraphs to summarize the temporal event data $\mathcal{X}$, versus using only a single hypergraph that aggregates all the events together. We therefore construct an \emph{inverse compression ratio} $\eta$ which computes our compression gain as 
\begin{align}\label{eq:compratio}
\eta = \frac{\mathcal{L}_{\text{MDL}}}{\mathcal{L}(K=1)},    
\end{align}
where $\mathcal{L}(K=1)$ is the description length of Eq.~\ref{eq:Ldecoupled} when all events are put into a single event cluster. A value $\eta=1$ implies that the event dataset $\mathcal{X}$ is not compressible using multiple hypergraphs, while $\eta\ll 1$ implies that the event dataset $\mathcal{X}$ can be greatly compressed using a representation of multiple hypergraphs.  

Evaluating $\mathcal{L}^{([i,j])}_{cluster}$ requires the evaluation of constant-time terms and two $\log\Omega$ terms which have nontrivial time complexity. Computing $\log\Omega(\bm{s}^{(k)},\bm{d}^{(k)})$ requires $\text{O}(S+D)$ operations once the margin counts $\bm{s}^{(k)},\bm{d}^{(k)}$ are known, and these margin counts require $\text{O}(m_k)$ operations to compute since we must iterate through the events in the cluster $C_k$. If the events are roughly evenly spaced in time, we would expect that $m_k\approx N(j-i)/T$ events would occur within the interval $[i,j]$, and the total complexity of evaluating the term $\log\Omega(\bm{s}^{(k)},\bm{d}^{(k)})$ would be $\text{O}(S+D+N(j-i)/T)$. Meanwhile, evaluating the term $\log\Omega(\mathcal{G}^{(k)},\bm{n}^{(k)})$ will have complexity $\text{O}(SD+\tau_k)$, with $\tau_k=j-i$ in this case. Combining these operations gives a total complexity of approximately $\text{O}(N(j-i)/T+SD+(j-i))$ for evaluating $\mathcal{L}^{([i,j])}_{cluster}$. Iterating over all $j$ and $i$ in the recursion then gives a total complexity of roughly $\text{O}((SD+N+T)T^2)$ for the dynamic programming algorithm using a na\"ive implementation. 

However, we can speed up the method by saving the $\mathcal{L}^{([i,j])}_{cluster}$ values as they're computed, requiring order $\text{O}(T^2)$ space. $\mathcal{L}^{([i,j])}_{cluster}$ can be computed from $\mathcal{L}^{([i-1,j])}_{cluster}$ in constant time if time step $i$ has no events. Similarly, $\mathcal{L}^{([i,j])}_{cluster}$ can be computed from $\mathcal{L}^{([i,j-1])}_{cluster}$ in constant time if time step $j$ has no events. Thus, when we loop over $j\in [1,T]$ and $i\in [1,j]$, we will only have to perform an $\text{O}(N(j-i)/T+SD+(j-i))$ evaluation of $\mathcal{L}^{([i,j])}_{cluster}$ when both cells $i$ and $j$ contain an event. Otherwise, we perform a constant-time update using $\mathcal{L}^{([i-1,j])}_{cluster}$ or $\mathcal{L}^{([i,j-1])}_{cluster}$ (depending on which endpoint does not have any event). 

If $T\gg N$ for $N$ constant---many temporal grid cells have no event---we have roughly $N$ unique cells containing events, resulting in ${N\choose 2}\ll {T\choose 2}$ pairs $\{i,j\}$ for which both the interval endpoints of $[i,j]$ contain an event, and for which we must perform the entire $\text{O}(N(j-i)/T+SD+(j-i))$ evaluation of $\mathcal{L}^{([i,j])}_{cluster}$. Thus, a vanishingly small fraction of pairs $\{i,j\}$ require these nontrivial evaluations, so the $\text{O}(T^2)$ evaluations will require roughly $\text{O}(T^2)$ total runtime to compute if $SD,N\ll T$ with $SD$ and $N$ constant with respect to $T$. 

When the above scaling assumptions for $SD$ and $N$ are not met, the $\text{O}(T^2)$ approximate total runtime of the dynamic programming method can break down. This is because the $\text{O}(N^2)$ nontrivial evaluations of complexity $\text{O}(N(j-i)/T+SD+(j-i))$ for $\mathcal{L}^{([i,j])}_{cluster}$ become important.  

Despite this speed-up, the time complexity of our exact dynamic programming solution may be too high for practical applications that involve large time periods or that require high values of $T$ for sufficient temporal resolution. In such cases we can use a greedy heuristic optimization method where we start with all time steps $i=1,...,T$ in their own cluster and iteratively merge the pair of time steps that gives the greatest decrease to the description length in Eq.~\ref{eq:Ldecoupled}. We save the description length changes induced by all proposed merges (including those that were sub-optimal) and perform greedy merges until all timesteps are in a single cluster. We then pick the value of $K$ for which the total description length was minimized over our set of merges. This greedy optimization method is not guaranteed to find the exact optimum, but it has a time complexity that is nearly $\text{O}(T)$ in practice, as for each $K=T,...,1$ we will only have to update the $\log \Omega$ terms for two merge pairs (those involving adjacent clusters to the one most recently merged). 

In the set of examples we have examined, this greedy method achieves MDL values that are nearly optimal but with much faster runtimes than the dynamic programming approach. See Sec.~\ref{sec:reconstruction} and Appendix~\ref{appendix:NYC} for numerical experiments computing the runtimes and inverse compression ratios achieved by the two algorithms on synthetic and real datasets respectively. However, one can never conclusively state that a greedy approximate method such as the one described above will be nearly optimal in a broader range of real applications without formal mathematical proof, which for the current situation is out of reach. See~\cite{aref2023heuristic} for a relevant discussion of the possible failures in approximate methods for optimizing the modularity objective for community detection in networks.

Code for the algorithms presented in this paper can be found at \url{https://github.com/aleckirkley/hypergraph-binning}.

%%%%%%%%%%%%%%%%%%%%%%%%%%%%%%%%%%%%%%%%%%%%%
%%%%%%%%%%%%%%%%%%%%%%%%%%%%%%%%%%%%%%%%%%%%%

\section{Results}
\label{sec:results}

\subsection{Reconstruction of synthetic data}
\label{sec:reconstruction}

To examine the performance of the algorithms presented in Sec.~\ref{sec:optimization}, we can generate synthetic data consisting of planted event clusters $\mathcal{C}$ with binnings $\bm{\tau}$ and test the ability for these algorithms to recover the planted clusters at various levels of injected noise. We generated synthetic datasets with $N\in [200,500,1000]$, $T\in [50,500]$, $K\in [2,5,10]$, and $S=D=5$ (the results did not depend on $S$ and $D$) in order to examine a range of model settings for the reconstruction tests. 

The synthetic event clusters $\mathcal{C}$ are generated by first drawing a partition of the $N$ events and $T$ timesteps into $K$ sets uniformly at random, then drawing the time step counts $\bm{n}^{(k)}$ uniformly at random within each cluster. To control the level of heterogeneity across the $K$ synthetic event clusters---which in turn controls the level of noise in the partition of the events, and consequently the reconstruction difficulty---our synthetic model includes a parameter $\gamma \geq 0$ which determines the localization of the edges $(s,d)$ within hypergraph $\bm{G}^{(k)}$ on sources $s\in \mathcal{S}$ and $d\in \mathcal{D}$. More specifically, for each cluster $C_k$ we independently generate the bipartite degrees $\bm{s}^{(k)}$ and $\bm{d}^{(k)}$ from a Dirichlet-Multinomial distribution with $m_k$ trials and concentration parameter $\gamma \bm{1}$, then draw the bipartite graph $\bm{G}^{(k)}$ at random from the set of non-negative integer matrices with row and column sums $\bm{s}^{(k)}$ and $\bm{d}^{(k)}$ using the algorithm of \cite{patefield1981algorithm}. This will create more localized bipartite degree distributions within hypergraph $\bm{G}^{(k)}$ (thus, more localized edge weights $G_{sd}^{(k)}$) and a higher variance across clusters as $\gamma \to 0$. The concentration parameter $\gamma$ thus serves as a tunable parameter that determines the distinguishability of the generated synthetic clusters, with $\gamma\to 0$ increasing the distinguishability of the clusters (e.g. increasing the signal-to-noise ratio).

\begin{figure}
    \centering
    \includegraphics[width=\columnwidth]{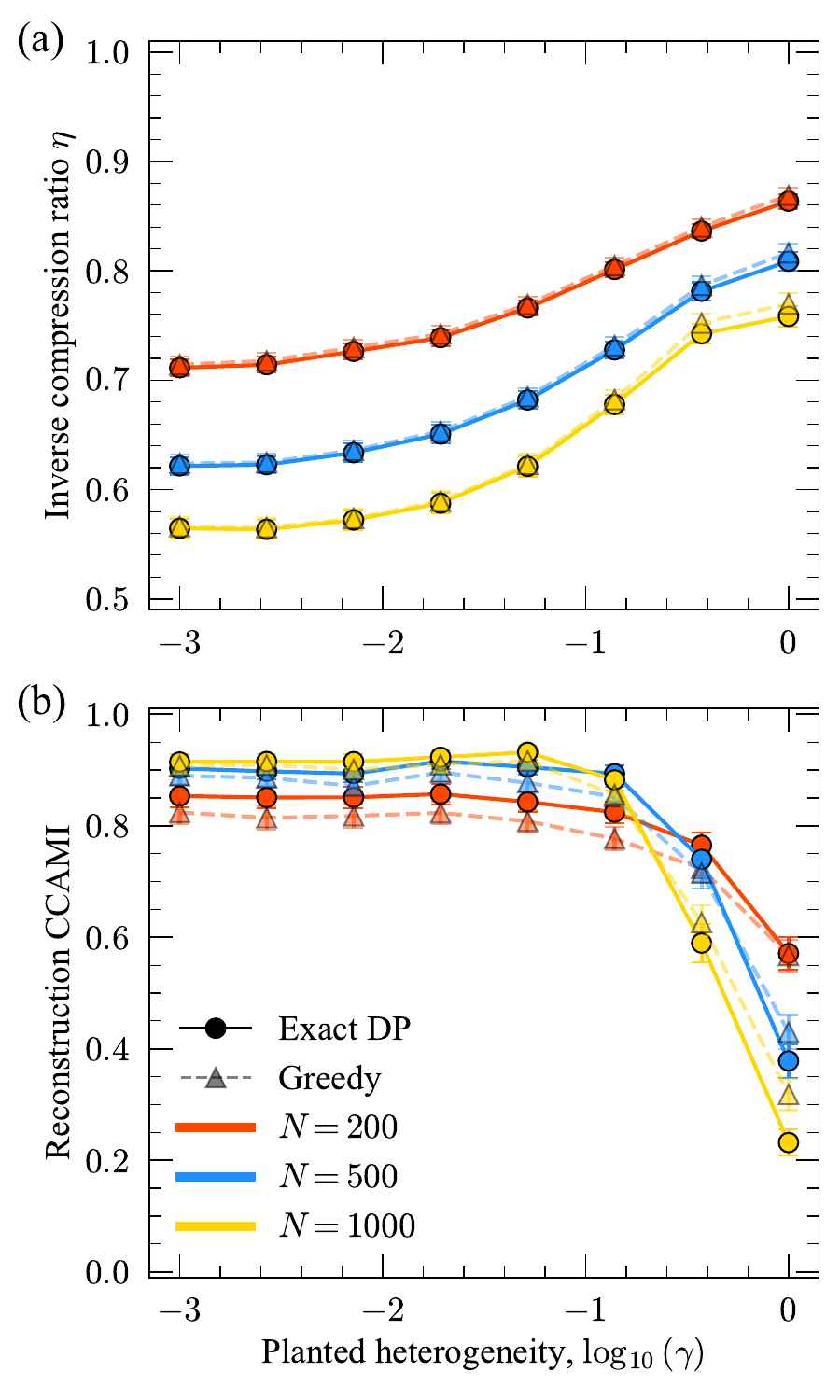}
    \caption{
    \textbf{Synthetic reconstruction performance.} \\
    \textbf{(a)} Average inverse compression ratio (Eq.~\ref{eq:compratio}) versus the logarithm of the planted level of cluster heterogeneity $\gamma$, for $N\in\{200,500,1000\}$ (line colors in red, blue, and yellow respectively). The exact dynamic programming algorithm results are shown with solid lines and circular markers, while the greedy algorithm results are shown with dotted lines and triangular markers. \textbf{(b)} Reconstruction accuracy, as quantified by the contiguity-corrected AMI (CCAMI, Eq.~\ref{eq:CCAMI}), over the same set of experiments. Averages for each panel are taken over 30 simulations with the parameters $\{S,D,K,T\}$ described in Sec.~\ref{sec:reconstruction}, and error bars represent 3 standard errors in the mean.  
    }
    \label{fig:reconstruction1}
\end{figure}

In Fig.~\ref{fig:reconstruction1} we plot the results of our synthetic reconstruction experiments. In Fig.~\ref{fig:reconstruction1}(a) we show the inverse compression ratio $\eta$ (Eq.~\ref{eq:compratio}) vs the logarithm of the planted heterogeneity $\gamma$ for $N\in [200,500,1000]$, averaged over 30 simulations at each combination of $K$ and $T$. Error bars represent three standard errors in the mean values estimated from the simulations, and the solid and dotted curves correspond to the dynamic programming and greedy algorithms described in Sec.~\ref{sec:optimization} respectively. We set $\Delta t=1$ for the reconstruction simulations. We can see that as the noise level $\gamma$ increases from $\gamma=10^{-3}$ to $\gamma=1$, the synthetic event clusters $\mathcal{X}$ become less and less compressible, and that more events $N$ results in better compression at all noise levels due to additional statistical evidence for the structure of each cluster. These results indicate that substantial data compression is possible using our algorithm, even at relatively high noise levels. We can also see very similar compression performance between the exact and greedy algorithms, indicating that, for these examples, the greedy method is achieving near-optimal compression at much lower computational cost than the dynamic programming method. 

As there are multiple binnings $\bm{\tau}$ that could correspond to any given set of event clusters $\mathcal{C}$---any binning that preserves the event partition while shifting the cluster boundaries in time---there is always a high level of uncertainty in reconstruction of $\bm{\tau}$, even with perfectly distinguishable event clusters $\mathcal{C}$. We therefore quantify the reconstruction accuracy in our simulations by representing an event partition $\mathcal{C}$ as a 1D partition of the temporally ordered event indices $\{1,...,N\}$, then compute a mutual information measure between the 1D partitions corresponding to the planted clusters $\mathcal{C}_{\text{pl}}$ and the clusters $\mathcal{C}_{\text{in}}$ inferred by our algorithm. However, standard mutual information-based measures~\cite{vinh2010information} are poorly suited for contiguous, low-dimensional partitions such as the ones we compare here, because they compute the partition similarity relative to an unconstrained (e.g., not necessarily contiguous) space of partitions that is much larger in size and much less structured than the space of contiguous partitions~\cite{kirkley2022spatial,kirkley2023compressing}. This results in artificially inflated values of the mutual information between contiguous partitions that may have little correlation other than that induced by their contiguity. 

With this in mind, here we construct a contiguity-corrected adjusted mutual information (CCAMI) to compute the similarity between the event clusters $\mathcal{C}_{\text{pl}}$ and $\mathcal{C}_{\text{in}}$, which is given by
\begin{align}\label{eq:CCAMI}
\text{CCAMI}(\mathcal{C}_{\text{pl}},\mathcal{C}_{\text{in}}) = \frac{\text{MI}(\mathcal{C}_{\text{pl}},\mathcal{C}_{\text{in}})-\langle 
\text{MI}(\mathcal{C}_{\text{pl}},\mathcal{C}_{\text{in}}) \rangle_{\text{c}}}{\text{max}\{\text{H}(\mathcal{C}_{\text{pl}}),\text{H}(\mathcal{C}_{\text{in}})\}-\langle 
\text{MI}(\mathcal{C}_{\text{pl}},\mathcal{C}_{\text{in}}) \rangle_{\text{c}}},    
\end{align}
where
\begin{align}
\text{MI}(\mathcal{C}_{\text{pl}},\mathcal{C}_{\text{in}}) = \text{H}(\mathcal{C}_{\text{pl}}) + \text{H}(\mathcal{C}_{\text{in}}) - \text{H}(\mathcal{C}_{\text{pl}},\mathcal{C}_{\text{in}})
\end{align}
is the standard Shannon mutual information between partitions $\mathcal{C}_{\text{pl}}$ and $\mathcal{C}_{\text{in}}$, $\text{H}(\mathcal{C})$ is the Shannon entropy of the cluster sizes in partition $\mathcal{C}$, and $\text{H}(\mathcal{C}_{\text{pl}},\mathcal{C}_{\text{in}})$ is the Shannon entropy of the joint partition defined by the overlap of the clusters in $\mathcal{C}_{\text{pl}}$ and $\mathcal{C}_{\text{in}}$. $\langle 
\text{MI}(\mathcal{C}_{\text{pl}},\mathcal{C}_{\text{in}}) \rangle_{\text{c}}$ is the expected value of this mutual information over all possible contiguous partitions with the same numbers of groups as $\mathcal{C}_{\text{in}}$ and $\mathcal{C}_{\text{pl}}$. Eq.~\ref{eq:CCAMI} quantifies how much information is shared between the planted event partition $\mathcal{C}_{\text{pl}}$ and our inferred event partition $\mathcal{C}_{\text{in}}$, relative to all pairs of contiguous partitions with the same numbers of clusters. In practice, $\langle 
\text{MI}(\mathcal{C}_{\text{pl}},\mathcal{C}_{\text{in}}) \rangle_{\text{c}}$ is difficult to compute analytically, so we estimate it using an average over 100 random draws of partitioning the $N$ events into $\abs{\mathcal{C}_{\text{in}}}$ and $\abs{\mathcal{C}_{\text{pl}}}$ contiguous clusters.  

In Fig.~\ref{fig:reconstruction1}(b) we show the reconstruction accuracy of our experiments as a function of the noise level $\gamma$, with the same parameter settings as in Fig.~\ref{fig:reconstruction1}(a). Consistent with the improved compression at lower $\gamma$ seen in Fig.~\ref{fig:reconstruction1}(a), we can see better reconstruction accuracy as $\gamma$ decreases and for synthetic datasets with a greater number of events $N$ when $\gamma \leq 10^{-1}$. When the noise level increases to $\gamma > 10^{-1}$, we observe a sharper drop in reconstruction accuracy for greater $N$, likely as a result of finite-size smoothing in phase transition-like behavior for the model detectability \cite{ricci2019typology}. We can also see that the exact and greedy algorithms have a non-negligible performance distinction with respect to reconstruction accuracy---as opposed to compression, as shown in Fig.~\ref{fig:reconstruction1}(a)---since in the low-noise regime the CCAMI values are noticeably lower for the greedy method than the exact method. This relative performance discrepancy for the greedy algorithm in Fig.~\ref{fig:reconstruction1}(a) and Fig.~\ref{fig:reconstruction1}(b) indicates that in some cases good data compression can be achieved for a variety of different partitions of the events in $\mathcal{X}$.

\begin{figure}
    \centering
   \hspace{-2em}
    \includegraphics[width=1.05\columnwidth]{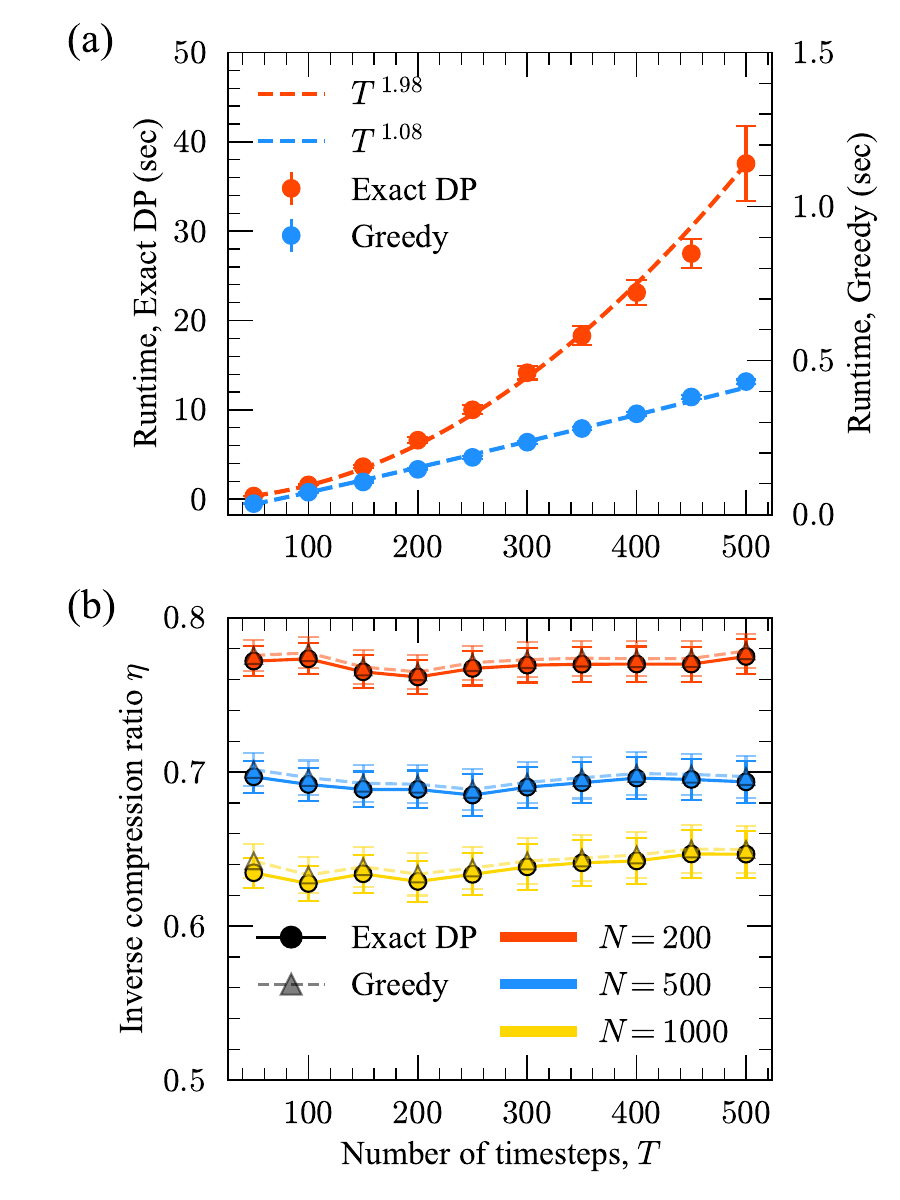}
    \caption{\textbf{Reconstruction parameter sensitivities.} \\
    \textbf{(a)} Average run time (in seconds) of reconstruction experiments (Fig.~\ref{fig:reconstruction1}) versus number of time steps $T$, for both algorithms described in Sec.~\ref{sec:optimization}. The performance of the exact dynamic programming algorithm is shown on the left axis, while that of the greedy algorithm is shown on the right axis. Regression lines of the form $\log (\text{Runtime})=\beta_1 \log (T) + \beta_2$, labeled with their least-squares estimates for the exponent $\hat \beta_1$, are shown as dotted lines. \textbf{(b)} Inverse compression ratio $\eta$ (Eq.~\ref{eq:compratio}) versus $T$ for the experiments conducted at different values of $N$. Averages for each panel are taken over 30 simulations with each combination of the parameters $\{S,D,K,\gamma\}$  described in Sec.~\ref{sec:reconstruction} (the averages in panel (a) also allow $N$ to vary), and error bars represent 3 standard errors in the mean.
    }
    \label{fig:reconstruction2}
\end{figure}

To verify the time complexities estimated in Sec.~\ref{sec:optimization}, in Fig.~\ref{fig:reconstruction2}(a) we plot the average run time for the dynamic programming algorithm (left axis) and greedy algorithm (right axis) as a function of the number of time steps $T$, for the reconstruction simulations in Fig.~\ref{fig:reconstruction1}. Along with these points we plot the regression lines obtained from fits of the form $\log (\text{Runtime})=\beta_1 \log (T) + \beta_2$, which are labelled with their least-squares estimates for the exponent $\hat \beta_1$. We can see that the dynamic program has a run time that scales approximately like $\text{O}(T^2)$, and that the greedy algorithm has a run time that is slightly worse than linear in the number of time steps $T$, while the absolute run times for the greedy algorithm are much faster than those for the dynamic program. 

In Fig.~\ref{fig:reconstruction2}(b), we plot the average inverse compression ratio $\eta$ of the reconstruction experiments as a function of the number of time steps $T$. We can see that the compression is essentially identical for all values of $T$, up to fluctuations. This indicates that the results of our algorithm are independent of the specific choice of temporal resolution $T$ as long as it is in a reasonable range (roughly at least on the order of the number of events $N$) that does not merge large time periods into single time steps.

\subsection{Foursquare checkins in NYC neighborhoods}
\label{sec:foursquare}

\begin{figure*}
    \centering
    \includegraphics[width=\textwidth]{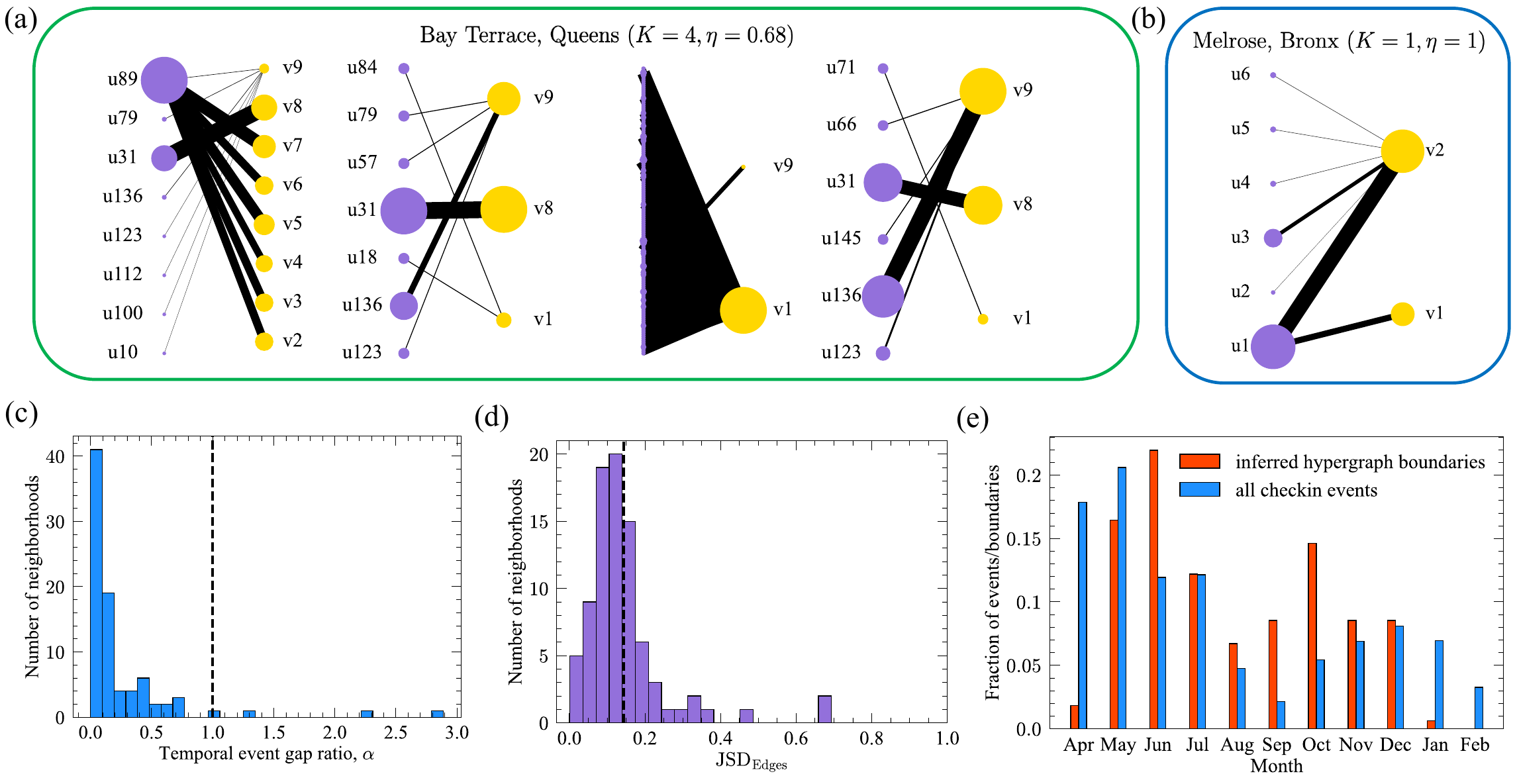}
    \caption{\textbf{FourSquare checkins in NYC neighborhoods.} The dataset, which aggregated checkins from April 2012 to February 2013 in New York City \cite{yang2014modeling,foursquareNYC}, consists of events $\bm{x}_i=(s_i,d_i,t_i)\in \mathcal{X}$ that denote a FourSquare checkin by a user $s_i$ at venue $d_i$ at time $t_i$. \textbf{(a)} Inferred hypergraphs for the Bay Terrace neighborhood, for which our method resulted in the highest level of data compression ($\eta = 0.68$). The hypergraphs are ordered chronologically left to right and shown in their incidence representation, with the width of edge $(s,d)$ proportional to the edge weight $G_{sd}^{(k)}$ which counts the number of events that contain user $s$ and venue $d$ in the time window. Source and destination nodes are scaled proportionally to their frequency of occurrence and labelled by unique user and venue ids respectively for each neighborhood. \textbf{(b)} Inferred hypergraph for Melrose, for which our method resulted in the lowest level of data compression ($\eta = 1$). \textbf{(c)} Histogram of the temporal event gap ratio $\alpha$ (Eq.~\ref{eq:alpha}) for all neighborhoods with $K>1$. \textbf{(d)} Histogram of the edge Jensen-Shannon Divergence $\text{JSD}_{\text{Edges}}$ (Eq.~\ref{eq:JSDnorm}) for all neighborhoods with $K>1$, with mean indicated using the dotted black line. \textbf{(e)} Fraction of all events (blue) and inferred temporal bin boundaries (red) that took place within each month, across all neighborhoods.}
    \label{fig:NYCNeigs}
\end{figure*}

To examine the performance of our method on real event data $\mathcal{X}$, we apply the algorithms of Sec.~\ref{sec:optimization} to a dataset of FourSquare checkins in New York City collected from April 2012 to February 2013 \cite{yang2014modeling,foursquareNYC}. In this dataset, each checkin event $\bm{x}_i=(s_i,d_i,t_i)\in \mathcal{X}$ denotes a FourSquare checkin by a user $s_i$ at venue $d_i$ at time $t_i$. Location-based social network (LBSN) data of this form are often used in urban planning, epidemiology, and sociology to understand human mobility co-location patterns \cite{chen2022contrasting,kadar2016exploring,antelmi2020design}, where users $s,s',s'',...\in \mathcal{S}$ are co-located if they check in at the same venue $d\in \mathcal{D}$ within some pre-defined time window. We can use the MDL method described in Sec.~\ref{sec:methods} to automatically extract a set of representative hypergraphs $\mathcal{G}$ from the LBSN checkin data $\mathcal{X}$ that capture homogeneous user activity patterns at different points in time. This allows us to, for instance, perform market segmentation to identify users $s\in \mathcal{S}$ with similar consumption patterns at different points in the year, or to identify seasonality in the congestion patterns at different venues.

To preprocess the FourSquare checkin data for analysis, we used neighborhood boundary shapefiles \cite{NYCneigs} to map the (latitude, longitude) pairs of the checkins to neighborhoods in NYC. We kept only the 1000 users and venues in the dataset with the most checkins and neighborhoods with at least 100 checkins over the 10 month period, with the aim of reducing biases and noise from users and venues with very infrequent app usage. The final dataset used in the analysis had $N=\text{64,366}$ events and $S=D=\text{1,000}$ users and venues spread across 91 neighborhoods. 

In Fig.~\ref{fig:NYCNeigs} we show the results of applying our exact dynamic programming method for hypergraph binning to the checkins for each neighborhood separately. This neighborhood-level analysis allows us to more easily visualize the inferred hypergraphs as well as perform cross-sectional comparisons across the neighborhoods regarding their event homogeneity, temporal burstiness, and compressibility. In our inference we set $\Delta t=$1 day. In Fig.~\ref{fig:NYCNeigs}(a) we show the inferred hypergraphs $\mathcal{G}=\{\bm{G}^{(1)},\bm{G}^{(2)},\bm{G}^{(3)},\bm{G}^{(4)}\}$ for the neighborhood (Bay Terrace, Queens) for which our method resulted in the highest level of data compression ($\eta = 0.68$). The hypergraphs $\bm{G}^{(k)}$, which are ordered chronologically left to right, are shown in their incidence representation, with the width of edge $(s,d)$ proportional to the edge weight $G_{sd}^{(k)}$ which counts the number of events that contain user $s$ and venue $d$ in the time period corresponding to hypergraph $\bm{G}^{(k)}$. Source and destination nodes in this representation are scaled in size proportionally to their weighted bipartite degrees (e.g. frequency of occurrence in events within the time period) and labelled by unique user and venue ids respectively for each neighborhood.   

We can see that the four inferred hypergraphs in Fig.~\ref{fig:NYCNeigs}(a) are very structurally distinct from one another. In the first time period inferred by our method (from the start of the study until July 9), corresponding to hypergraph $\bm{G}^{(1)}$, we observe that a large portion of the activity was dominated by user `u89', who visited venues 'v2' through `v7' as well as `v9', the last of which was also visited by the rest of the users except `u31'. User `u31' also made a substantial number of checkins during this period, but only to venue `v8'. The inclusion of these two distinct activity patterns (checkins by `u89' and `u31') in the hypergraph $\bm{G}^{(1)}$ is a result of both users performing their checkins consistently over the time period corresponding to the first hypergraph. In the second hypergraph (corresponding to the time period July 10 - October 26), we can see that user `u31' is still making consistent checkins at venue `v8', but that user `u89' is no longer making checkins. There is also turnover in the other users and venues. In the third hypergraph (corresponding to October 27 - November 1) we see a very different checkin activity pattern, which corresponds to the tropical storm Hurricane Sandy hitting New York City. Here we see many users (too many to be labelled in the figure) checking in at location `v1', Throgs Neck Bridge between Queens and the Bronx, likely signalling evacuation and return to the city. In the fourth hypergraph (corresponding to November 2 through the end of the study period), we see a return to normal with a somewhat similar activity structure as in the second hypergraph, where most checkins are performed by users `u31' and `u136' at venues `v8' and `v9' respectively. The checkin data for this neighborhood is easily compressed using our method due to these four very distinct periods of high localization of the events onto a few users and venues. 

In contrast, in Fig.~\ref{fig:NYCNeigs}(b), we see a very different story for Melrose in the Bronx. Here we see that the event data was optimally compressed into a single hypergraph (i.e. $K=\eta=1$), for which checkin activity is dominated by user `u1' and venue `v2'. (Note that these are not the same as user `u1' and venue `v2' in Bay Terrace, since these abbreviated user and venue IDs were generated separately for the two neighborhoods in the figure.) The neighborhood-level set of events for Melrose is incompressible using multiple hypergraphs, since it does not have multiple distinct periods of activity, instead exhibiting consistent checkins by one user at one venue. 

To quantify the extent of temporal localization in the hypergraphs inferred with our method, we define a temporal event gap ratio $\alpha$ as the ratio of the median inter-event time within clusters to the median inter-event time between clusters, or
\begin{align}\label{eq:alpha}
\alpha = \frac{\text{median}(\{t_{i+1}-t_{i}\vert c_i=c_{i+1}\}_{i=1}^{N-1})}{\text{median}(\{t_{i+1}-t_{i}\vert c_i\neq c_{i+1}\}_{i=1}^{N-1})},    
\end{align}
where $c_i\in \{1,...,K\}$ is the event cluster index of the $i$-th event $\bm{x}_i$. $\alpha < 1$ when the events within the event clusters tend to be more localized in time than the events on the borders of the clusters, and $\alpha > 1$ when the opposite is true. In Fig.~\ref{fig:NYCNeigs}(c), we plot a histogram of the ratio $\alpha$ for all neighborhoods analyzed that had an inferred $K>1$. We can see that the inferred hypergraphs in all but 4 neighborhoods had events that were more temporally localized than the pairs of events that transitioned between hypergraphs ($\alpha < 1$), indicating that our method is identifying periods of temporally localized activity in the LBSN data. 

\begin{figure*}
    \centering
    \includegraphics[width=\textwidth]{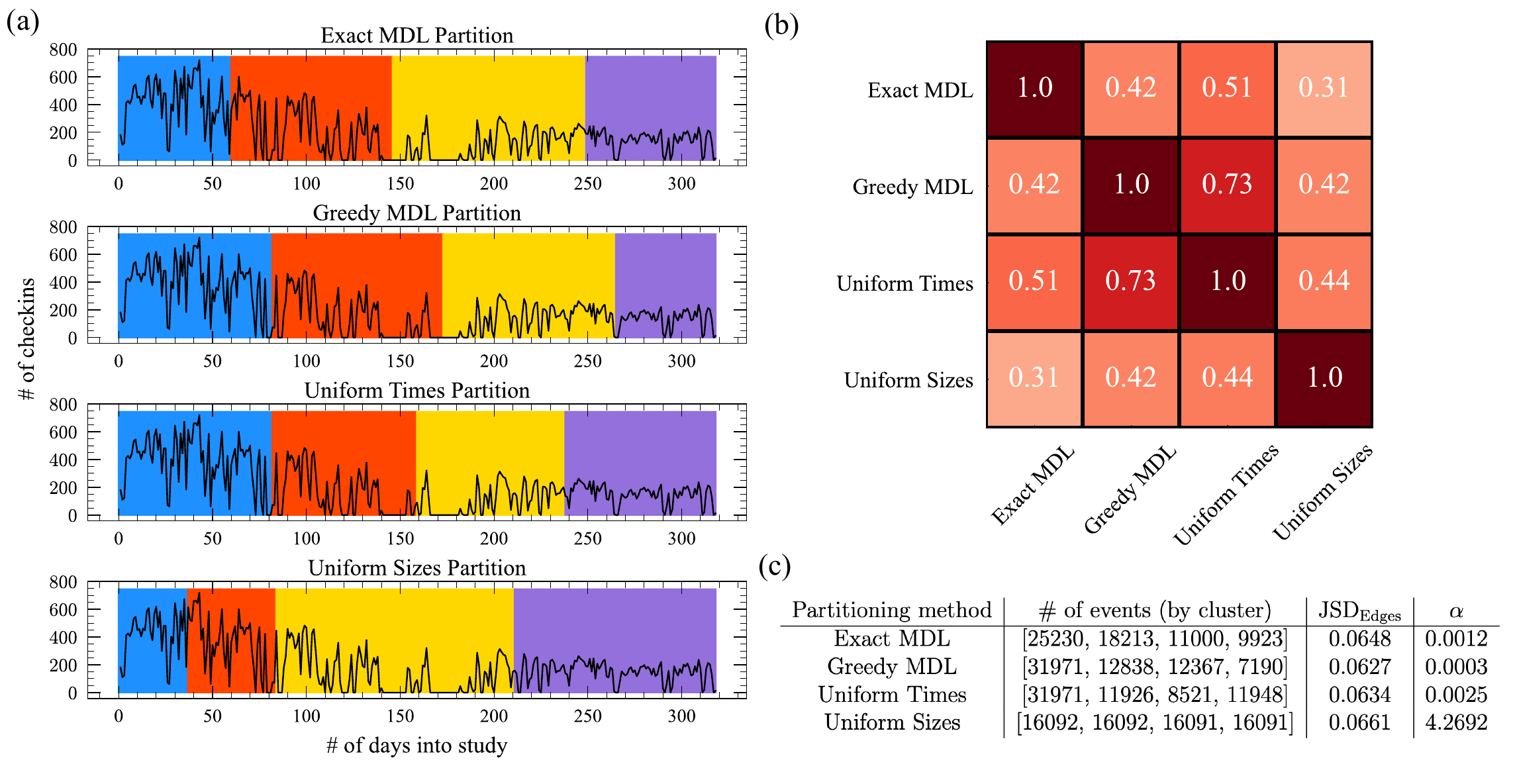}
    \caption{\textbf{FourSquare checkins across all of NYC.} \textbf{(a)} Binnings obtained when applying the exact dynamic programming method (top plot) and greedy agglomerative method (second plot) of Sec.~\ref{sec:optimization} to the set of checkins aggregated across all neighborhoods in NYC, with the number of checkins for each day of the study plotted as a solid black line underneath. Colors distinguish the $K=4$ different temporal bins inferred by each of these algorithms. The bottom two plots show the partitions obtained by na\"ively partitioning the events into $K=4$ time windows of equal duration and into time windows with an equal number of events (third and fourth plots respectively). \textbf{(b)} CCAMI matrix among all pairs of the four partitions shown in panel (a). \textbf{(c)} Table of summary statistics for the partitions in panel (a).}
    \label{fig:NYCFull}
\end{figure*}

To examine the localization of inferred hypergraphs on sources and destinations relative to the overall localization in the dataset $\mathcal{X}$, we compute the expected reduction in uncertainty for predicting the source and destination $(s,d)$ of a randomly chosen edge in $\mathcal{G}$ versus the fully aggregated hypergraph $\bm{G}_0=\bigcup_{k=1}^{K}\bm{G}^{(k)}$. If the edges $(s,d)$ in each hypergraph $\bm{G}^{(k)}$ are much more highly localized on source/destination pairs than in the overall data $\mathcal{X}$, it is substantially easier to predict the label $(s,d)$ of a randomly chosen edge in $\mathcal{G}$ than in $\bm{G}_0$. The reduction in our predictive uncertainty in going from $\bm{G}_0$ to $\mathcal{G}$ can be quantified by the Generalized Jensen-Shannon Divergence~\cite{nielsen2020generalization,kirkley2020information}, given by
\begin{align}\label{eq:JSDunnorm}
\text{JSD}_{\text{Edges, unnormalized}} =  \text{H}(\bm{G}_0) - \sum_{k=1}^{K}\frac{m_k}{N}\text{H}(\bm{G}^{(k)}),
\end{align}
where
\begin{align}
\text{H}(\bm{G}_0) &= -\sum_{s,d=1}^{S,D}\left(\frac{\sum_{k=1}^{K}G^{(k)}_{sd}}{N}\right)\log \left(\frac{\sum_{k=1}^{K}G^{(k)}_{sd}}{N}\right), \\ 
\text{H}(\bm{G}^{(k)}) &= -\sum_{s,d=1}^{S,D}\left(\frac{G^{(k)}_{sd}}{m_k}\right)\log \left(\frac{G^{(k)}_{sd}}{m_k}\right)    
\end{align}
are the Shannon entropies of the edges in the aggregated graph and inferred hypergraphs respectively. One can show that Eq.~\ref{eq:JSDunnorm} is bounded below by $0$ (due to the concavity of entropy) and above by $H(\bm{G}_0)$, so we can rescale the JSD to $[0,1]$ by dividing by this upper bound, thus
\begin{align}\label{eq:JSDnorm}
\text{JSD}_{\text{Edges}} = 1 - \frac{1}{\text{H}(\bm{G}_0)}\sum_{k=1}^{K}\frac{m_k}{N}\text{H}(\bm{G}^{(k)}).    
\end{align}
Eq.~\ref{eq:JSDnorm} tells us the fraction of information (in terms of predictive power for the edge labels) we lose by using the aggregated hypergraph $\bm{G}_0$ instead of the cluster-level hypergraphs $\mathcal{G}=\{\bm{G}^{(k)}\}_{k=1}^{K}$. A value of $\text{JSD}_{\text{Edges}}\approx 0$ indicates little edge localization within the clusters relative to the overall dataset, and a value of $\text{JSD}_{\text{Edges}}\approx 1$ indicates high edge localization within the clusters relative to the overall dataset. 

In Fig.~\ref{fig:NYCNeigs}(d), we plot a histogram of $\text{JSD}_{\text{Edges}}$ for all neighborhoods analyzed that had an inferred $K>1$, with the distribution mean of $\text{JSD}_{\text{Edges}}\approx 0.15$ indicated with the vertical black line. The mean JSD value of $0.15$ indicates a relatively high average level of localization among the edges $(s,d)$ within each hypergraph. We can also observe that all but 5 neighborhoods had an information gain of at least $5\%$ relative to the overall data $\mathcal{X}$, indicating non-negligible localization of the sources/destinations in the hypergraphs inferred with our method.  

Finally, in Fig.~\ref{fig:NYCNeigs}(e) we plot a histogram showing the fraction of all events that took place within each month (blue) and the fraction of inferred temporal snapshot boundaries that took place within each month (red). Here we can observe substantial differences in these distributions, indicating that the inferred boundaries are to some extent negatively correlated with the temporal density of events. For example, there is a sizable drop in event frequency from May to June, and we see a spike in the number of inferred temporal boundaries in June, suggesting that the drop in events provided sufficient statistical evidence for the formation of a new cluster boundary in the information theoretically optimal binnings. We also see large discrepancies for September and October, where there are comparatively many boundaries but few events. This may be correlated with the uptick in the overall density of events, seasonal fluctuations in consumer behavior, and Hurricane Sandy (for October).

In Appendix~\ref{appendix:NYC}, we run additional tests using the neighborhood-level event data, in order to understand the discrepancies between the exact dynamic programming algorithm and the fast greedy agglomerative algorithm of Sec.~\ref{sec:optimization} when applied to this dataset.

One can also examine the fluctuations in large-scale checkin patterns at the level of the entire city by aggregating the events over all neighborhoods. In the top two panels of Fig.~\ref{fig:NYCFull}(a), we show the binnings obtained when applying our exact dynamic programming method and greedy agglomerative method to the aggregated dataset representing the top 1000 users and venues in the FourSquare dataset across all neighborhoods. Different colors distinguish the $K=4$ different temporal bins obtained  by each of these algorithms, and the number of checkins for each day of the study is plotted as a solid black line. 

In the bottom two plots of Fig.~\ref{fig:NYCFull}(a), we show the partitions obtained using na\"ive binning heuristics with the same number of clusters ($K=4$)---a partition of the events into time windows of equal duration (third plot), and a partition of the events into time windows with an equal number of events (fourth plot). We can see substantial shifts in the cluster boundaries for these methods relative to the MDL-based methods, despite the (relatively strong) constraint of fixing the value of $K$. The MDL-based methods have inferred boundaries in the gaps with sparse event data around days 145 (exact method) and 162 (greedy method), due to their focus on temporally localized clusters of events, while the other methods have boundaries that are uncorrelated with temporal event density. However, a binning heuristic that only looks for temporal gaps in events will also fail to reproduce the partitions obtained by the MDL methods, since we see some sizable gaps included within the inferred clusterings of both methods, where there is enough statistical evidence for the algorithm to create a contiguous cluster due to localization of the events on sources and destinations. 

In Fig.~\ref{fig:NYCFull}(b), we plot a matrix of the CCAMI values between each pair of partitions among the four shown in Fig.~\ref{fig:NYCFull}(a). We observe that, despite the apparent visual similarity of some pairs of partitions, the information shared between many pairs is only modestly more than what one would expect in random partitions of the time interval into $K=4$ clusters. We also can see that---despite having a description length ($\eta_{greedy}=0.758$) which is comparable to the description length of the optimal partition obtained by the exact dynamic programming algorithm ($\eta_{exact}=0.750$)---the greedy partition is actually quite different than the true MDL-optimal partition when considering the strong constraints imposed by contiguity ($\text{CCAMI}=0.42$). In fact, the greedy MDL partition is less similar to the true MDL-optimal partition than the partition obtained by simply splitting the interval into $K=4$ windows of equal duration. This highlights the importance of our exact dynamic programming solution, and is consistent with findings in network community detection that identify high levels of degeneracy in the near-optimal partitions of networks  \cite{GDC10,peixoto2021revealing,kirkley2022representative,aref2023heuristic}. 

We can also compute the inverse compression ratio of Eq.~\ref{eq:compratio} for the baseline partitions by plugging these partitions directly into the objective in Eq.~\ref{eq:Ltotal}. We find inverse compression ratios of $0.764$ and $0.778$ respectively for the partitions whose clusters are uniform in time and the number of events respectively, which correspond to states that are roughly $2^{600}$ and $2^{1800}$ times worse than the greedy solution (which is $2^{700}$ times worse than the exact optimum) in terms of relative posterior probability.  

In Fig.~\ref{fig:NYCFull}(c), we plot summary statistics of the inferred hypergraphs using each binning method. Mirroring Fig.~\ref{fig:NYCFull}(a) we can see high variability in the number of events within the clusters across the four partitions. We can also see that the exact MDL approach has the best balance of edge localization ($\text{JSD}_{\text{Edges}}=0.0648$) and temporal localization ($\alpha=0.0012$) among its inferred event clusters. While it is only the second best method regarding each metric individually---e.g., it has the second-highest $\text{JSD}_{\text{Edges}}$ value and the second lowest $\alpha$ value---the top performers in $\text{JSD}_{\text{Edges}}$ (Uniform Sizes) and $\alpha$ (Greedy MDL) are the worst performers regarding $\alpha$ and $\text{JSD}_{\text{Edges}}$ respectively.

%%%%%%%%%%%%%%%%%%%%%%%%%%%%%%%%%%%%%%%%%%%%%
%%%%%%%%%%%%%%%%%%%%%%%%%%%%%%%%%%%%%%%%%%%%%

\section{Conclusion}

In this paper we develop a nonparametric approach for inferring representative hypergraph snapshots from temporal event data based on the MDL principle. Our approach considers the problem of transmitting the data to a receiver in multiple stages of increasing granularity, with the hypergraph snapshots as an intermediate step. The configuration of hypergraphs that minimizes the description length of this transmission process is then selected as the MDL-optimal hypergraph representation of the data. Our method automatically performs model selection for the number and composition of the hypergraphs with no parameter tuning. We employ an exact dynamic programming algorithm to identify the hypergraphs that minimize our description length objective in a runtime that scales quadratically with the number of discrete time steps in the limit of high temporal resolution. We also develop a fast greedy agglomerative algorithm that achieves near-optimal configurations with substantially reduced run times in the examples studied, enabling the application of our method to large datasets. We demonstrate that our methods are able to consistently reconstruct synthetic data with planted hypergraph structure even with appreciable noise, and can reveal meaningful representative structures in real location-based social network data to understand human mobility patterns. 

There are a number of ways our methods can be extended in future work. In this paper we explore a data encoding that exploits redundancy provided by degree heterogeneity in the incidence representations of the representative hypergraphs within the data, but one can in principle exploit other structure as well to develop efficient encodings. In a Bayesian framing of the hypergraph inference problem, these alternative encodings would correspond to generative models other than the configuration-style model corresponding to the encoding of this paper. Alternative structure that could be potentially exploited for improved compression may include community structure, transitivity, or overlaps among hyperedges. One can also impose asymmetry in the encoding between the source and destination nodes to more clearly highlight the desired hypergraph structure, or incorporate other relevant metadata on the edges such as weights or temporal duration of events. Finally, there is no guarantee that the greedy method for minimizing the description length will be near-optimal in all problem settings, so a more comprehensive evaluation of this method in other applications or a mathematical proof of its approximation capabilities is important for future work.

%%%%%%%%%%%%%%%%%%%%%%%%%%%%%%%%%%%%%%%%%%%%%
%%%%%%%%%%%%%%%%%%%%%%%%%%%%%%%%%%%%%%%%%%%%%

\section*{Acknowledgments}
\vspace{-\baselineskip}
This research was supported in part by the HKU-100 Start Up Grant and the HKU Institute of  Data Science Research Seed Fund. The author thanks Shihui Feng for helpful discussions.

%%%%%%%%%%%%%%%%%%%%%%%%%%%%%%%%%%%%%%%%%%%%%
%%%%%%%%%%%%%%%%%%%%%%%%%%%%%%%%%%%%%%%%%%%%%
% \bibliographystyle{numeric}
% \bibliography{refs}

\begin{thebibliography}{10}
\expandafter\ifx\csname url\endcsname\relax
  \def\url#1{\texttt{#1}}\fi
\expandafter\ifx\csname urlprefix\endcsname\relax\def\urlprefix{URL }\fi

\bibitem{battiston2020networks}
F.~Battiston, G.~Cencetti, I.~Iacopini, V.~Latora, M.~Lucas, A.~Patania, J.-G.
  Young, and G.~Petri, Networks beyond pairwise interactions: Structure and
  dynamics. \textit{Physics Reports} \textbf{874}, 1--92 (2020).

\bibitem{battiston2021physics}
F.~Battiston, E.~Amico, A.~Barrat, G.~Bianconi, G.~Ferraz~de Arruda,
  B.~Franceschiello, I.~Iacopini, S.~K{\'e}fi, V.~Latora, Y.~Moreno,
  \textit{et~al.}, The physics of higher-order interactions in complex systems.
  \textit{Nature Physics} \textbf{17}(10), 1093--1098 (2021).

\bibitem{li2022spatial}
Z.~Li, C.~Huang, L.~Xia, Y.~Xu, and J.~Pei, Spatial-temporal hypergraph
  self-supervised learning for crime prediction. In \textit{2022 IEEE 38th
  International Conference on Data Engineering (ICDE)}, pp. 2984--2996, IEEE
  (2022).

\bibitem{proferes2021studying}
N.~Proferes, N.~Jones, S.~Gilbert, C.~Fiesler, and M.~Zimmer, Studying reddit:
  A systematic overview of disciplines, approaches, methods, and ethics.
  \textit{Social Media + Society} \textbf{7}(2), 20563051211019004 (2021).

\bibitem{antelmi2020design}
A.~Antelmi, G.~Cordasco, C.~Spagnuolo, and V.~Scarano, A design-methodology for
  epidemic dynamics via time-varying hypergraphs. In \textit{Proceedings of the
  19th International Conference on Autonomous Agents and MultiAgent Systems},
  pp. 61--69 (2020).

\bibitem{kim2015product}
Y.~Kim and R.~Krishnan, On product-level uncertainty and online purchase
  behavior: An empirical analysis. \textit{Management Science} \textbf{61}(10),
  2449--2467 (2015).

\bibitem{eren2020multi}
M.~E. Eren, J.~S. Moore, and B.~S. Alexandro, Multi-dimensional anomalous
  entity detection via Poisson tensor factorization. In \textit{2020 IEEE
  International Conference on Intelligence and Security Informatics (ISI)}, pp.
  1--6, IEEE (2020).

\bibitem{huang2017unsupervised}
C.~Huang, D.~Wang, and J.~Tao, An unsupervised approach to inferring the
  localness of people using incomplete geotemporal online check-in data.
  \textit{ACM Transactions on Intelligent Systems and Technology (TIST)}
  \textbf{8}(6), 1--18 (2017).

\bibitem{balmaki2022modern}
B.~Balmaki, M.~A. Rostami, T.~Christensen, E.~A. Leger, J.~M. Allen, C.~R.
  Feldman, M.~L. Forister, and L.~A. Dyer, Modern approaches for leveraging
  biodiversity collections to understand change in plant-insect interactions.
  \textit{Frontiers in Ecology and Evolution} \textbf{10}, 924941 (2022).

\bibitem{holme2012temporal}
P.~Holme and J.~Saram{\"a}ki, Temporal networks. \textit{Physics Reports}
  \textbf{519}(3), 97--125 (2012).

\bibitem{cencetti2021temporal}
G.~Cencetti, F.~Battiston, B.~Lepri, and M.~Karsai, Temporal properties of
  higher-order interactions in social networks. \textit{Scientific Reports}
  \textbf{11}(1), 7028 (2021).

\bibitem{myers2023topological}
A.~Myers, C.~Joslyn, B.~Kay, E.~Purvine, G.~Roek, and M.~Shapiro, Topological
  analysis of temporal hypergraphs. In \textit{Algorithms and Models for the
  Web Graph: 18th International Workshop, WAW 2023, Toronto, ON, Canada, May
  23--26, 2023, Proceedings}, pp. 127--146, Springer (2023).

\bibitem{lee2023temporal}
G.~Lee and K.~Shin, Temporal hypergraph motifs. \textit{Knowledge and
  Information Systems} \textbf{65}(4), 1549--1586 (2023).

\bibitem{taylor2019supracentrality}
D.~Taylor, M.~A. Porter, and P.~J. Mucha, Supracentrality analysis of temporal
  networks with directed interlayer coupling. \textit{Temporal Network Theory}
  pp. 325--344 (2019).

\bibitem{amburg2020clustering}
I.~Amburg, N.~Veldt, and A.~Benson, Clustering in graphs and hypergraphs with
  categorical edge labels. In \textit{Proceedings of The Web Conference 2020},
  pp. 706--717 (2020).

\bibitem{huang2020temporal}
S.~Huang, Z.~Bao, G.~Li, Y.~Zhou, and J.~S. Culpepper, Temporal network
  representation learning via historical neighborhoods aggregation. In
  \textit{2020 IEEE 36th International Conference on Data Engineering (ICDE)},
  pp. 1117--1128, IEEE (2020).

\bibitem{neuhauser2021consensus}
L.~Neuh{\"a}user, R.~Lambiotte, and M.~T. Schaub, Consensus dynamics on
  temporal hypergraphs. \textit{Physical Review E} \textbf{104}(6), 064305
  (2021).

\bibitem{li2017fundamental}
A.~Li, S.~P. Cornelius, Y.-Y. Liu, L.~Wang, and A.-L. Barab{\'a}si, The
  fundamental advantages of temporal networks. \textit{Science}
  \textbf{358}(6366), 1042--1046 (2017).

\bibitem{valdano2015analytical}
E.~Valdano, L.~Ferreri, C.~Poletto, and V.~Colizza, Analytical computation of
  the epidemic threshold on temporal networks. \textit{Physical Review X}
  \textbf{5}(2), 021005 (2015).

\bibitem{schwarz2020temporal}
B.~Schwarz, D.~P. V{\'a}zquez, P.~J. CaraDonna, T.~M. Knight, G.~Benadi, C.~F.
  Dormann, B.~Gauzens, E.~Motivans, J.~Resasco, N.~Bl{\"u}thgen,
  \textit{et~al.}, Temporal scale-dependence of plant--pollinator networks.
  \textit{Oikos} \textbf{129}(9), 1289--1302 (2020).

\bibitem{nicosia2013graph}
V.~Nicosia, J.~Tang, C.~Mascolo, M.~Musolesi, G.~Russo, and V.~Latora, Graph
  metrics for temporal networks. \textit{Temporal Networks} pp. 15--40 (2013).

\bibitem{zha2016unfolding}
Y.~Zha, T.~Zhou, and C.~Zhou, Unfolding large-scale online collaborative human
  dynamics. \textit{Proceedings of the National Academy of Sciences}
  \textbf{113}(51), 14627--14632 (2016).

\bibitem{rissanen1978}
J.~Rissanen, Modeling by the shortest data description. \textit{Automatica}
  \textbf{14}, 465--471 (1978).

\bibitem{grunwald2007minimum}
P.~D. Gr{\"u}nwald and A.~Gr{\"u}nwald, \textit{The Minimum Description Length
  Principle}. MIT Press, Cambridge, MA (2007).

\bibitem{Rosvall07}
M.~Rosvall and C.~T. Bergstrom, An information-theoretic framework for
  resolving community structure in complex networks. \textit{Proceedings of the
  National Academy of Sciences} \textbf{104}, 7327--7331 (2007).

\bibitem{Peixoto14a}
T.~P. Peixoto, Hierarchical block structures and high-resolution model
  selection in large networks. \textit{Physical Review X} \textbf{4}, 011047
  (2014).

\bibitem{Kirkley22Reps}
A.~Kirkley and M.~E.~J. Newman, Representative community divisions of networks.
  \textit{Communications Physics} \textbf{5}, 40 (2022).

\bibitem{koutra2014vog}
D.~Koutra, U.~Kang, J.~Vreeken, and C.~Faloutsos, Vog: Summarizing and
  understanding large graphs. In \textit{Proceedings of the 2014 SIAM
  international conference on data mining}, pp. 91--99, SIAM (2014).

\bibitem{wegner2014subgraph}
A.~E. Wegner, Subgraph covers: an information-theoretic approach to motif
  analysis in networks. \textit{Physical Review X} \textbf{4}(4), 041026
  (2014).

\bibitem{bloem2020large}
P.~Bloem and S.~de~Rooij, Large-scale network motif analysis using compression.
  \textit{Data Mining and Knowledge Discovery} \textbf{34}(5), 1421--1453
  (2020).

\bibitem{young2021hypergraph}
J.-G. Young, G.~Petri, and T.~P. Peixoto, Hypergraph reconstruction from
  network data. \textit{Communications Physics} \textbf{4}(1), 1--11 (2021).

\bibitem{bouritsas2021partition}
G.~Bouritsas, A.~Loukas, N.~Karalias, and M.~Bronstein, Partition and code:
  learning how to compress graphs. \textit{Advances in Neural Information
  Processing Systems} \textbf{34}, 18603--18619 (2021).

\bibitem{feng2012summarization}
J.~Feng, X.~He, B.~Konte, C.~B{\"o}hm, and C.~Plant, Summarization-based mining
  bipartite graphs. In \textit{Proceedings of the 18th ACM SIGKDD international
  conference on Knowledge discovery and data mining}, pp. 1249--1257 (2012).

\bibitem{zhou2021dpgs}
H.~Zhou, S.~Liu, K.~Lee, K.~Shin, H.~Shen, and X.~Cheng, Dpgs:
  Degree-preserving graph summarization. In \textit{Proceedings of the 2021
  SIAM International Conference on Data Mining (SDM)}, pp. 280--288, SIAM
  (2021).

\bibitem{koutra2015summarizing}
D.~Koutra, U.~Kang, J.~Vreeken, and C.~Faloutsos, Summarizing and understanding
  large graphs. \textit{Statistical Analysis and Data Mining: The ASA Data
  Science Journal} \textbf{8}(3), 183--202 (2015).

\bibitem{masuda2019detecting}
N.~Masuda and P.~Holme, Detecting sequences of system states in temporal
  networks. \textit{Scientific Reports} \textbf{9}(1), 1--11 (2019).

\bibitem{de2015structural}
M.~De~Domenico, V.~Nicosia, A.~Arenas, and V.~Latora, Structural reducibility
  of multilayer networks. \textit{Nature {C}ommunications} \textbf{6}(1), 1--9
  (2015).

\bibitem{kirkley2023compressing}
A.~Kirkley, A.~Rojas, M.~Rosvall, and J.-G. Young, Compressing network
  populations with modal networks reveals structural diversity.
  \textit{Communications Physics} \textbf{6}, 148 (2022).

\bibitem{gurukar2022leveraging}
S.~Gurukar, B.~Boettner, C.~Browning, C.~Calder, and S.~Parthasarathy,
  Leveraging network representation learning and community detection for
  analyzing the activity profiles of adolescents. \textit{Applied Network
  Science} \textbf{7}(1), 27 (2022).

\bibitem{yu2016spatial}
W.~Yu, Spatial co-location pattern mining for location-based services in road
  networks. \textit{Expert Systems with Applications} \textbf{46}, 324--335
  (2016).

\bibitem{zheng2018novel}
X.~Zheng, Y.~Luo, L.~Sun, X.~Ding, and J.~Zhang, A novel social network hybrid
  recommender system based on hypergraph topologic structure. \textit{World
  Wide Web} \textbf{21}, 985--1013 (2018).

\bibitem{burgos2008two}
E.~Burgos, H.~Ceva, L.~Hern{\'a}ndez, R.~P. Perazzo, M.~Devoto, and D.~Medan,
  Two classes of bipartite networks: nested biological and social systems.
  \textit{Physical Review E} \textbf{78}(4), 046113 (2008).

\bibitem{lee2021hyperedges}
G.~Lee, M.~Choe, and K.~Shin, How do hyperedges overlap in real-world
  hypergraphs?-patterns, measures, and generators. In \textit{Proceedings of
  the Web Conference 2021}, pp. 3396--3407 (2021).

\bibitem{zhang2010hypergraph}
Z.-K. Zhang and C.~Liu, A hypergraph model of social tagging networks.
  \textit{Journal of Statistical Mechanics: Theory and Experiment}
  \textbf{2010}(10), P10005 (2010).

\bibitem{feller1950}
W.~Feller, \textit{An Introduction to Probability Theory and Its Applications}. Wiley, Hoboken, NJ (1950).

\bibitem{jerdee2022improved}
M.~Jerdee, A.~Kirkley, and M.~Newman, Improved estimates for the number of
  non-negative integer matrices with given row and column sums. \textit{Proceedings of the Royal Society
  A} \textbf{480}(20230470) (2024).

\bibitem{jackson2005algorithm}
B.~Jackson, J.~D. Scargle, D.~Barnes, S.~Arabhi, A.~Alt, P.~Gioumousis,
  E.~Gwin, P.~Sangtrakulcharoen, L.~Tan, and T.~T. Tsai, An algorithm for
  optimal partitioning of data on an interval. \textit{IEEE Signal Processing
  Letters} \textbf{12}(2), 105--108 (2005).

\bibitem{bellman2013dynamic}
R.~Bellman, \textit{Dynamic Programming}. Princeton University Press (1957).

\bibitem{patania2023rapid}
A.~Patania, A.~Allard, and J.-G. Young, Exact and rapid linear clustering of
  networks with dynamic programming. \textit{Proceedings of the Royal Society
  A} \textbf{479}(2275) (2023).

\bibitem{aref2023modularity}
S.~Aref, M.~Mostajabdaveh, and H.~Chheda, Heuristic Modularity Maximization Algorithms for Community Detection Rarely Return an Optimal Partition or Anything Similar. \textit{arXiv preprint arXiv:2302.14698} (2023).


\bibitem{patefield1981algorithm}
W.~Patefield, Algorithm as 159: an efficient method of generating random
  r $\times$ c tables with given row and column totals. \textit{Journal of the
  Royal Statistical Society. Series C (Applied Statistics)} \textbf{30}(1),
  91--97 (1981).

\bibitem{vinh2010information}
N.~X. Vinh, J.~Epps, and J.~Bailey, Information theoretic measures for
  clusterings comparison: Variants, properties, normalization and correction
  for chance. \textit{The Journal of Machine Learning Research} \textbf{11},
  2837--2854 (2010).

\bibitem{kirkley2022spatial}
A.~Kirkley, Spatial regionalization based on optimal information compression.
  \textit{Communications Physics} \textbf{5}(1), 249 (2022).

\bibitem{ricci2019typology}
F.~Ricci-Tersenghi, G.~Semerjian, and L.~Zdeborov{\'a}, Typology of phase
  transitions in bayesian inference problems. \textit{Physical Review E}
  \textbf{99}(4), 042109 (2019).

\bibitem{yang2014modeling}
D.~Yang, D.~Zhang, V.~W. Zheng, and Z.~Yu, Modeling user activity preference by
  leveraging user spatial temporal characteristics in lbsns. \textit{IEEE
  Transactions on Systems, Man, and Cybernetics: Systems} \textbf{45}(1),
  129--142 (2014).

\bibitem{foursquareNYC}
Foursquare - NYC and Tokyo check-ins.
  \url{https://www.kaggle.com/datasets/chetanism/foursquare-nyc-and-tokyo-checkin-dataset?resource=download}.
  Accessed: 2023-06-15.

\bibitem{chen2022contrasting}
Z.~Chen, S.~Kelty, A.~G. Evsukoff, B.~F. Welles, J.~Bagrow, R.~Menezes, and
  G.~Ghoshal, Contrasting social and non-social sources of predictability in
  human mobility. \textit{Nature Communications} \textbf{13}(1), 1922 (2022).

\bibitem{kadar2016exploring}
C.~Kadar, J.~Iria, and I.~P. Cvijikj, Exploring foursquare-derived features for
  crime prediction in new york city. \textit{KDD-Urban Computing WS}
  \textbf{16}, 10--1145 (2016).

\bibitem{NYCneigs}
Pediacities-NYC-neighborhoods.
  \url{https://data.beta.nyc/dataset/pediacities-nyc-neighborhoods/resource/35dd04fb-81b3-479b-a074-a27a37888ce7}.
  Accessed: 2023-06-15.

\bibitem{nielsen2020generalization}
F.~Nielsen, On a generalization of the Jensen--Shannon divergence and the
  jensen--shannon centroid. \textit{Entropy} \textbf{22}(2), 221 (2020).

\bibitem{kirkley2020information}
A.~Kirkley, Information theoretic network approach to socioeconomic
  correlations. \textit{Physical Review Research} \textbf{2}, 043212 (2020).

\bibitem{GDC10}
B.~H. Good, Y.-A. de~Montjoye, and A.~Clauset, Performance of modularity
  maximization in practical contexts. \textit{Physical Review E} \textbf{81},
  046106 (2010).

\bibitem{peixoto2021revealing}
T.~P. Peixoto, Revealing consensus and dissensus between network partitions.
  \textit{Physical Review X} \textbf{11}, 021003 (2021).

\bibitem{kirkley2022representative}
A.~Kirkley and M.~Newman, Representative community divisions of networks.
  \textit{Communications Physics} \textbf{5}(1), 40 (2022).

\bibitem{aref2023heuristic}
S.~Aref, M.~Mostajabdaveh, and H.~Chheda, Heuristic modularity maximization
  algorithms for community detection rarely return an optimal partition or
  anything similar. \textit{arXiv preprint arXiv:2302.14698}  (2023).


  
  
  

\end{thebibliography}

%%%%%%%%%%%%%%%%%%%%%%%%%%%%%%%%%%%%%%%%%%%%%
%%%%%%%%%%%%%%%%%%%%%%%%%%%%%%%%%%%%%%%%%%%%%

\clearpage
\appendix
\onecolumngrid

\section{Additional Tests with NYC Checkins Dataset}
\label{appendix:NYC}

In this Appendix we plot the results of a variety of tests to compare the performance of the exact dynamic programming algorithm and greedy algorithm of Sec.~\ref{sec:optimization} when applied to the FourSquare checkins dataset of Sec.~\ref{sec:foursquare}. We also report summary statistics of the 91 neighborhoods studied.

\begin{figure}[b]
    \centering
    \includegraphics[width=0.6\textwidth]{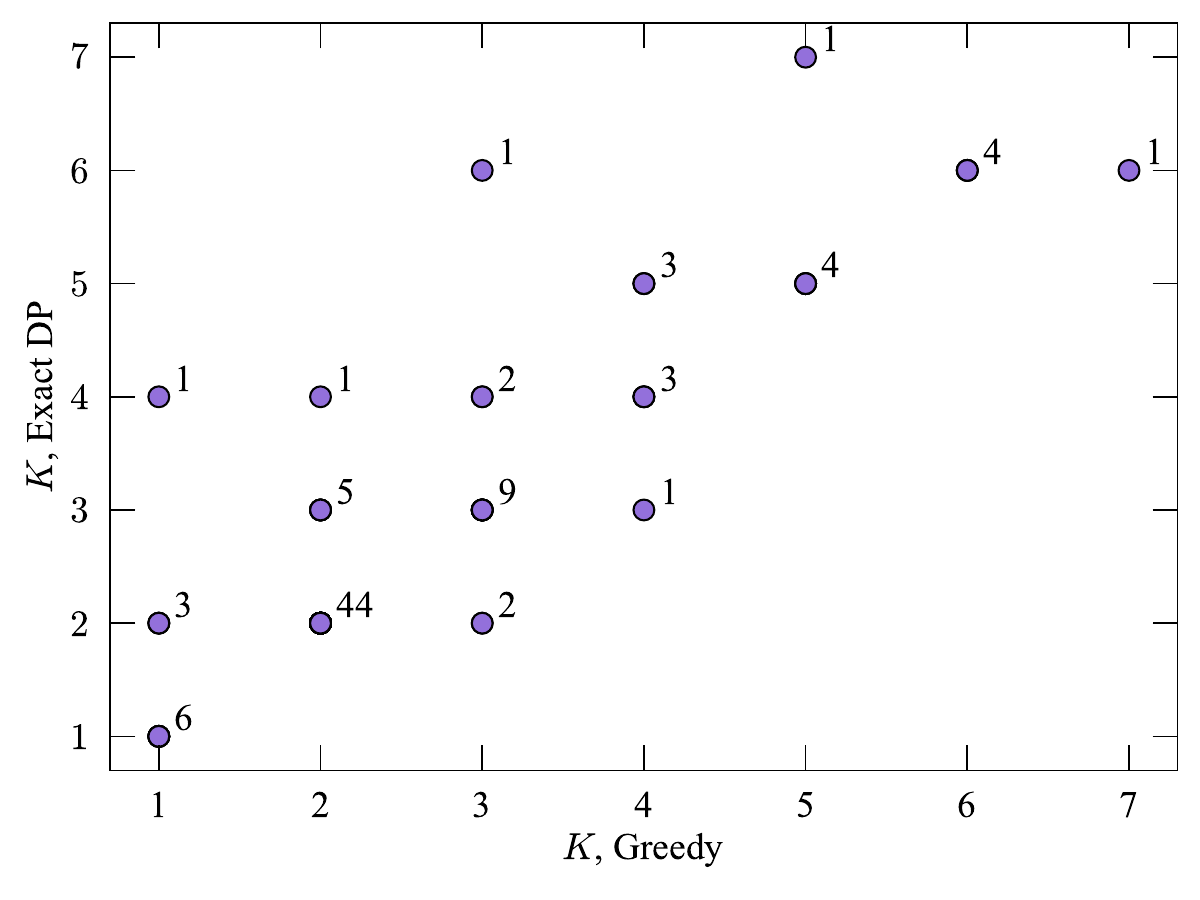}
    \caption{Number of clusters inferred by the exact dynamic programming algorithm (y-axis) and the greedy algorithm (x-axis). Data point labels indicate the number of neighborhoods with the given combination $(K_{greedy},K_{exact})$.}
    \label{fig:NYCKcounts}
\end{figure}

\begin{figure}
    \centering
    \includegraphics[width=0.6\textwidth]{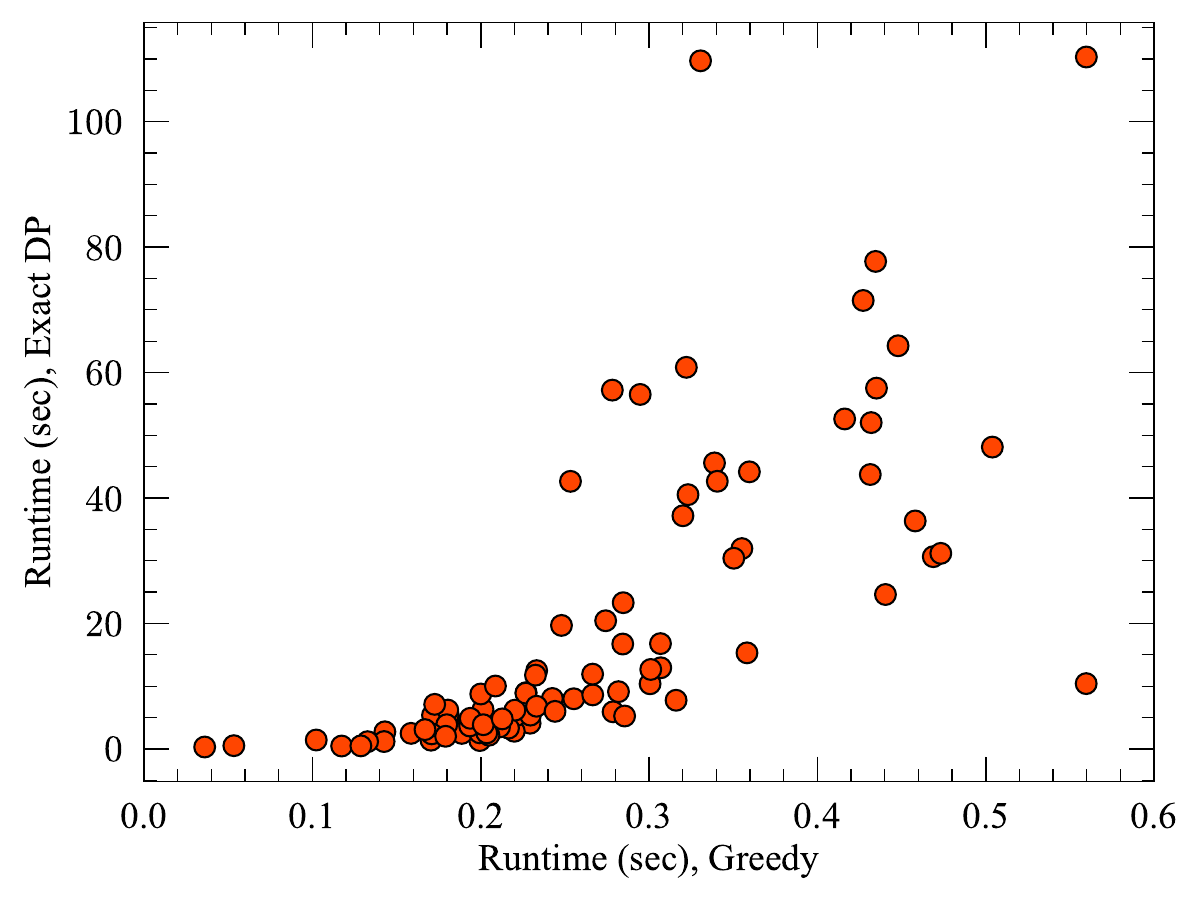}
    \caption{Run time of the exact dynamic programming algorithm (y-axis) and the greedy algorithm (x-axis) for each neighborhood in the study.}
    \label{fig:NYCruntimes}
\end{figure}

\begin{figure}
    \centering
    \includegraphics[width=0.6\textwidth]{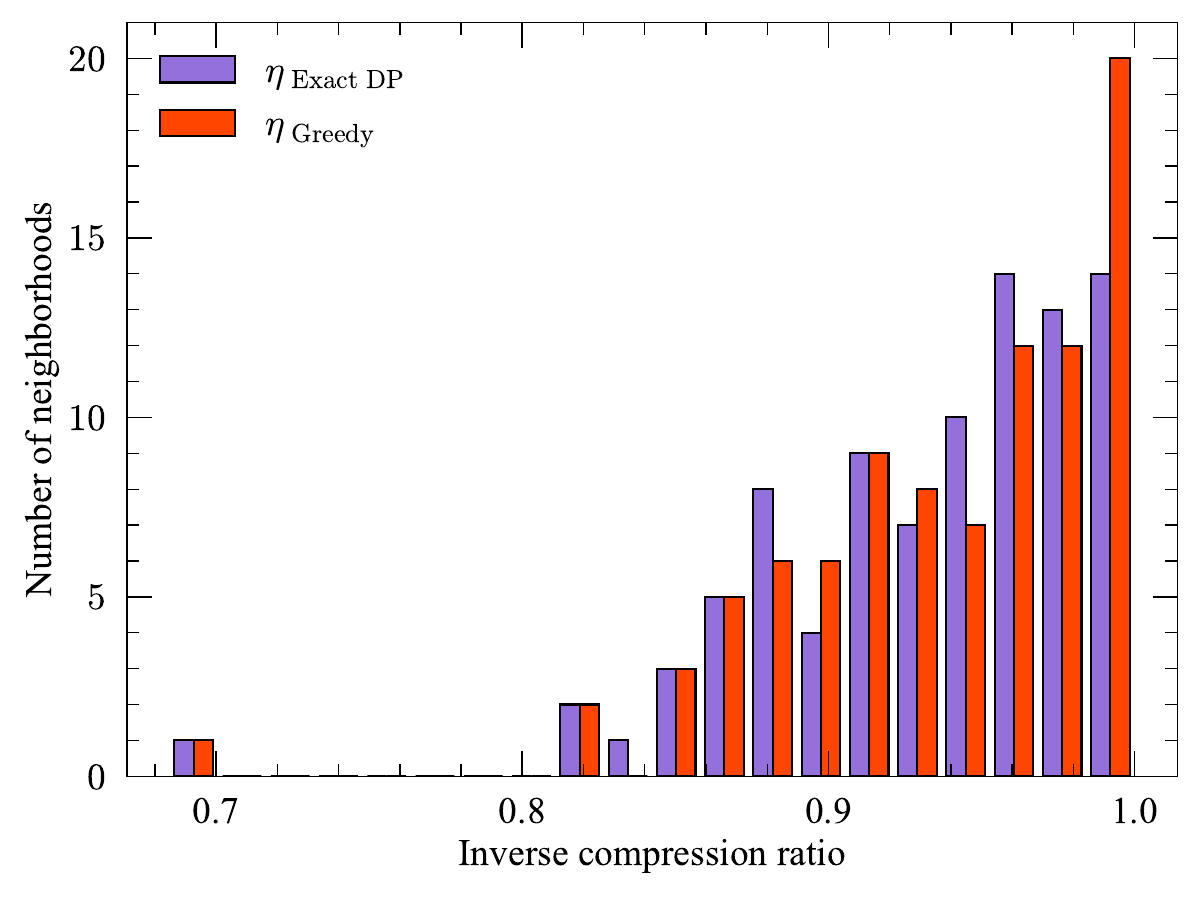}
    \caption{Histograms of the compression ratio $\eta$ (Eq.~\ref{eq:compratio}) for the event data in each neighborhood in the study, after compression with the exact method (purple) and greedy method (red).}
    \label{fig:NYCcomps}
\end{figure}

\begin{figure}
    \centering
    \includegraphics[width=0.6\textwidth]{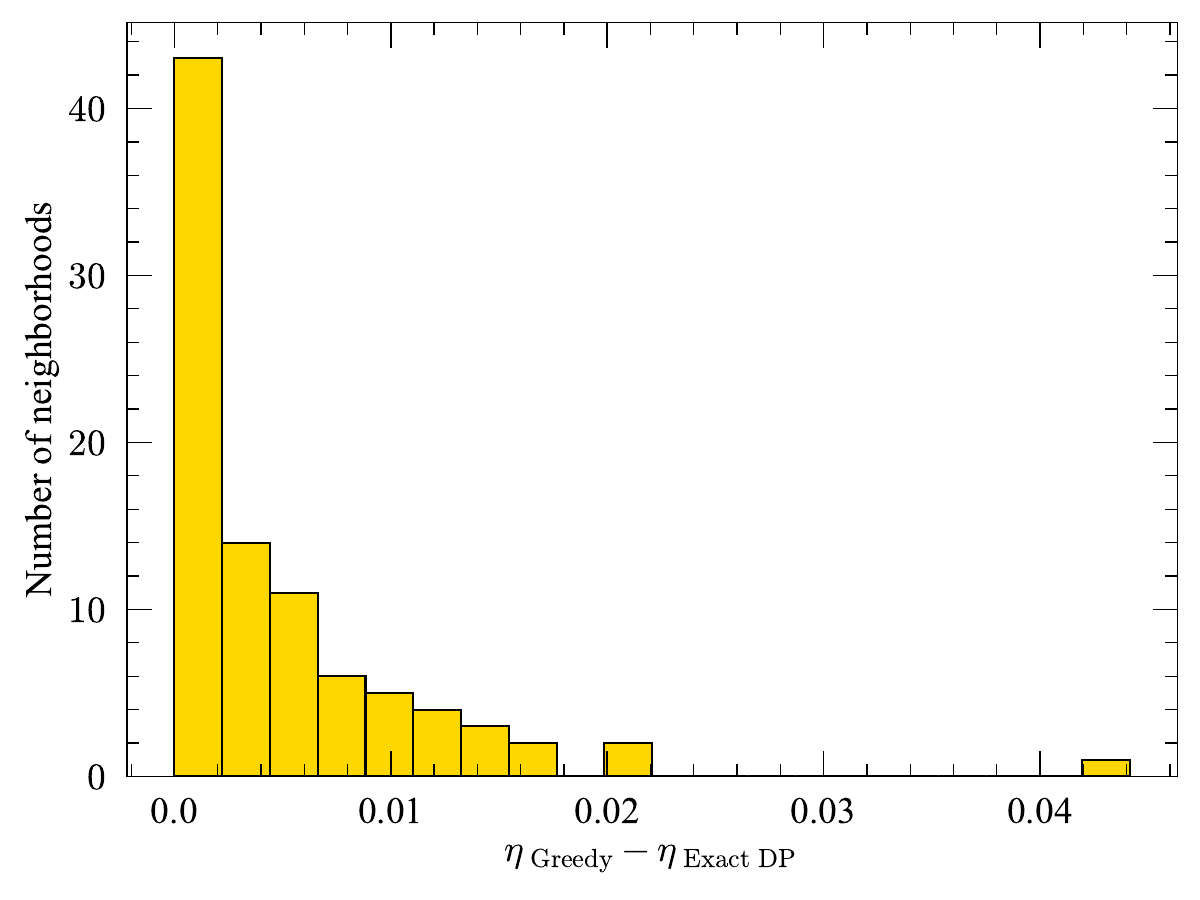}
    \caption{Histograms of the differences in the compression ratio $\eta$ (Eq.~\ref{eq:compratio}) between the exact method and greedy method, for each neighborhood in the study.}
    \label{fig:NYCcompdiffs}
\end{figure}

\begin{figure}
    \centering
    \includegraphics[width=0.6\textwidth]{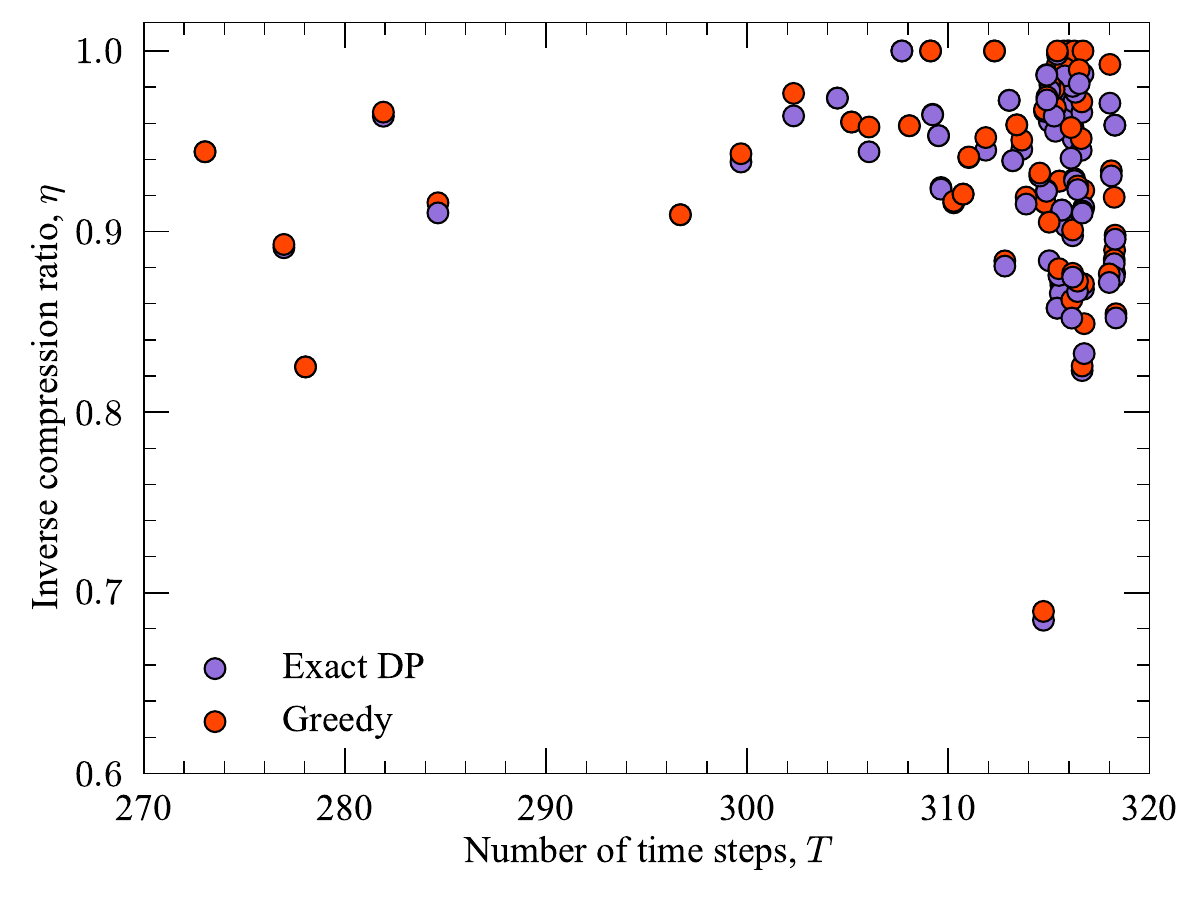}
    \caption{Inverse compression ratio $\eta$ versus number of time steps $T$ for each neighborhood in the study.}
    \label{fig:NYCcompVsT}
\end{figure}

\begin{figure}
    \centering
    \includegraphics[width=0.6\textwidth]{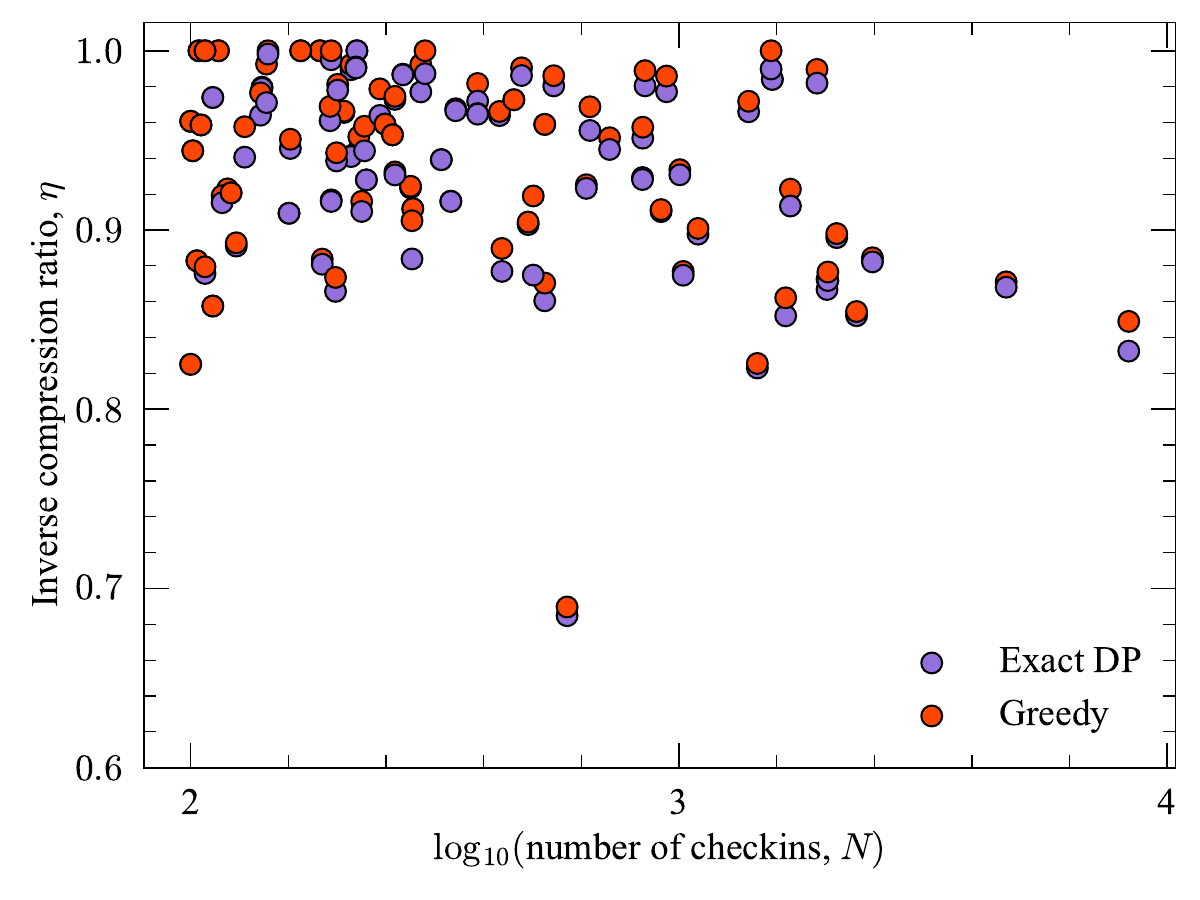}
    \caption{Inverse compression ratio $\eta$ versus the logarithm of the number of checkins $N$ for each neighborhood in the study.}
    \label{fig:NYCcompVsN}
\end{figure}

\begin{table}
\centering
\begin{tabular}{|c|c|c|c|c|}
\hline
\# &                          ~~~~Neighborhood~~~~ &~~~~$N$~~~~&~~~~$S$~~~~&~~~~$D$~~~~\\ 
 \hline

0  &                                Arverne &   103 &    1 &    2 \\ 
 \hline
1  &                                Astoria &   852 &   65 &   16 \\ 
 \hline
2  &                      Battery Park City &   184 &   38 &    2 \\ 
 \hline
3  &                            Bay Terrace &   590 &  145 &    9 \\ 
 \hline
4  &                     Bedford-Stuyvesant &  1553 &   31 &   21 \\ 
 \hline
5  &                            Bensonhurst &   186 &   11 &    3 \\ 
 \hline
6  &                            Boerum Hill &   159 &    7 &    3 \\ 
 \hline
7  &                           Borough Park &   213 &   14 &    4 \\ 
 \hline
8  &                         Brighton Beach &   429 &   63 &    4 \\ 
 \hline
9  &                              Bronxdale &   100 &    9 &    2 \\ 
 \hline
10 &                       Brooklyn Heights &   206 &    9 &    3 \\ 
 \hline
11 &                            Brownsville &   434 &   14 &    7 \\ 
 \hline
12 &                            Bull's Head &   229 &    6 &    6 \\ 
 \hline
13 &                               Canarsie &   491 &   13 &    9 \\ 
 \hline
14 &                           Central Park &   531 &  189 &    8 \\ 
 \hline
15 &                                Chelsea &  4678 &  654 &   60 \\ 
 \hline
16 &                           Civic Center &   219 &   26 &    3 \\ 
 \hline
17 &                             Co-op City &   124 &   12 &    2 \\ 
 \hline
18 &                          College Point &   459 &   28 &    9 \\ 
 \hline
19 &                              Concourse &   341 &  169 &    2 \\ 
 \hline
20 &                      Concourse Village &   194 &   24 &    4 \\ 
 \hline
21 &                           Coney Island &   114 &    5 &    2 \\ 
 \hline
22 &                           Country Club &   140 &    2 &    2 \\ 
 \hline
23 &                          Crown Heights &   101 &    8 &    3 \\ 
 \hline
24 &                          Cypress Hills &   282 &   25 &    3 \\ 
 \hline
25 &                                  DUMBO &   119 &   21 &    3 \\ 
 \hline
26 &                       Ditmars Steinway &   296 &    7 &    4 \\ 
 \hline
27 &                             Douglaston &   193 &    5 &    3 \\ 
 \hline
28 &                      Downtown Brooklyn &   387 &   47 &    8 \\ 
 \hline
29 &                          East Elmhurst &   198 &    6 &    5 \\ 
 \hline
30 &                          East Flatbush &   224 &   40 &    1 \\ 
 \hline
31 &                            East Harlem &   843 &   62 &   14 \\ 
 \hline
32 &                          East New York &  1389 &    6 &   19 \\ 
 \hline
33 &                           East Village &  1691 &  398 &   30 \\ 
 \hline
34 &                               Elmhurst &   218 &   48 &    4 \\ 
 \hline
35 &                     Financial District &  1447 &  218 &   23 \\ 
 \hline
36 &                               Flatbush &   349 &   21 &    7 \\ 
 \hline
37 &                      Flatiron District &  2310 &  298 &   27 \\ 
 \hline
38 &                               Flushing &   721 &   40 &   14 \\ 
 \hline
39 &           Flushing Meadows Corona Park &   221 &   90 &    3 \\ 
 \hline
40 &                                Fordham &   139 &   15 &    3 \\ 
 \hline
41 &                           Forest Hills &   213 &   22 &    4 \\ 
 \hline
42 &                            Fort Greene &   227 &   83 &    4 \\ 
 \hline
43 &                                Gowanus &   194 &   28 &    3 \\ 
 \hline
44 &                               Gramercy &  1094 &  266 &   12 \\ 
 \hline
45 &                              Gravesend &   285 &   12 &    5 \\ 
 \hline
\end{tabular}
\caption{Neighborhood-level checkin dataset size details (0 - 45).}
\end{table}

\begin{table}
\centering
\begin{tabular}{|c|c|c|c|c|}
\hline
\# &                          ~~~~Neighborhood~~~~ &~~~~$N$~~~~&~~~~$S$~~~~&~~~~$D$~~~~\\ 
 \hline

46 &                             Greenpoint &   111 &   29 &    2 \\ 
 \hline
47 &                      Greenwich Village &   842 &  252 &   16 \\ 
 \hline
48 &                                 Harlem &  1544 &   76 &   27 \\ 
 \hline
49 &                         Hell's Kitchen &  2490 &  411 &   37 \\ 
 \hline
50 &                                 Inwood &   262 &   13 &    5 \\ 
 \hline
51 &                                Jamaica &   657 &   84 &    7 \\ 
 \hline
52 &                        Jamaica Estates &   105 &    9 &    2 \\ 
 \hline
53 &  John F. Kennedy International Airport &   919 &  315 &    4 \\ 
 \hline
54 &                               Kips Bay &   503 &  163 &    9 \\ 
 \hline
55 &                      LaGuardia Airport &   646 &  259 &    3 \\ 
 \hline
56 &                       Long Island City &   943 &   68 &   16 \\ 
 \hline
57 &                               Longwood &   284 &    9 &    8 \\ 
 \hline
58 &                        Lower East Side &   554 &  107 &   11 \\ 
 \hline
59 &                                Melrose &   168 &    6 &    2 \\ 
 \hline
60 &                         Middle Village &   104 &    1 &    1 \\ 
 \hline
61 &                                Midtown &  8338 &  687 &  126 \\ 
 \hline
62 &                                Midwood &   107 &    5 &    2 \\ 
 \hline
63 &                             Mill Basin &   107 &   21 &    3 \\ 
 \hline
64 &                    Morningside Heights &   116 &   12 &    2 \\ 
 \hline
65 &                            Murray Hill &   476 &   18 &   11 \\ 
 \hline
66 &                                   NoHo &   244 &   39 &    5 \\ 
 \hline
67 &                                Norwood &   200 &   13 &    3 \\ 
 \hline
68 &                             Pelham Bay &   100 &    9 &    2 \\ 
 \hline
69 &                       Prospect Heights &   129 &   73 &    2 \\ 
 \hline
70 &                          Prospect Park &   272 &   71 &    3 \\ 
 \hline
71 &                         Queens Village &   111 &    2 &    2 \\ 
 \hline
72 &                          Richmond Hill &   121 &    8 &    2 \\ 
 \hline
73 &                       Roosevelt Island &   160 &   39 &    2 \\ 
 \hline
74 &                         Sheepshead Bay &   531 &   57 &   10 \\ 
 \hline
75 &                                   SoHo &  1654 &  181 &   29 \\ 
 \hline
76 &                       South Ozone Park &   387 &    3 &    6 \\ 
 \hline
77 &                            South Slope &   143 &    8 &    3 \\ 
 \hline
78 &                             St. George &   199 &   37 &    2 \\ 
 \hline
79 &                              Stapleton &   144 &    3 &    2 \\ 
 \hline
80 &                            Sunset Park &   326 &   16 &    8 \\ 
 \hline
81 &                       Theater District &  2104 &  415 &   33 \\ 
 \hline
82 &                                Tribeca &   302 &   27 &    7 \\ 
 \hline
83 &                              Unionport &   250 &   12 &    4 \\ 
 \hline
84 &                        Upper East Side &  2010 &  264 &   37 \\ 
 \hline
85 &                        Upper West Side &  2018 &  283 &   35 \\ 
 \hline
86 &                     Washington Heights &  1004 &   28 &   16 \\ 
 \hline
87 &                           West Village &  1020 &  199 &   22 \\ 
 \hline
88 &                             Whitestone &   259 &    5 &    6 \\ 
 \hline
89 &                           Williamsburg &  1917 &  139 &   31 \\ 
 \hline
90 &                               Woodside &   262 &   49 &    5 \\ 
 \hline
\end{tabular}
\caption{Neighborhood-level checkin dataset size details (46 - 90).}
\end{table}

\end{document}